\documentclass[a4paper,fleqn]{cas-dc}

\usepackage[authoryear]{natbib}
\usepackage{bbm}
\usepackage{bm}
\usepackage{amsfonts}
\usepackage{multirow}
\usepackage{caption}
\usepackage{subcaption}
\usepackage{array}
\usepackage{threeparttable}
\usepackage{newtxtext}

\def\tsc#1{\csdef{#1}{\textsc{\lowercase{#1}}\xspace}}
\tsc{WGM}
\tsc{QE}

\begin{document}
\let\WriteBookmarks\relax
\def\floatpagepagefraction{1}
\def\textpagefraction{.001}

\shorttitle{Olfactory memory in \emph{Drosophila}}    

\shortauthors{C. Rohlfs}  

\title [mode = title]{A descriptive analysis of olfactory sensation and memory in \emph{Drosophila} and its relation to artificial neural networks}  



%

\author{Chris Rohlfs}[orcid=0000-0001-7714-9231]



\ead{car2228@columbia.edu}



\affiliation{organization={Columbia University Department of Electrical Engineering},
            addressline={Mudd 1310, 500 West $120^{th}$ Street}, 
            city={New York},
            postcode={10027-6623}, 
            state={NY},
            country={USA}}


\begin{abstract}
This article provides a background and descriptive analysis of insect memory and the coding of olfactory sensation in \emph{Drosophila}, presenting graphs and summary statistics from a large dataset of neurons and synapses that was recently made publicly available and also discussing findings from the existing empirical literature. Some general principles from \emph{Drosophila} olfaction are discussed as they apply to the design of analogous systems in artificial neural networks: (1) the networks used for coding are shallow; (2) the level of connectedness varies widely across neurons in the same layer; (3) much communication is between neurons in the same layer; (4) in most olfactory learning, the manner in which sensory inputs are represented in stored memory is largely fixed, and the learning process involves developing positive or negative associations with existing categories of inputs.
\end{abstract}


\maketitle

\section{Introduction} \label{Olfactory Introduction}

Memory is a key element of cognitive functioning that involves distinct neural pathways and regions of the brain \citep{Fernandes2020, Hamm2019, Matsumoto2018, Thompson1996}. Both the committing and the retrieval of information from memory are highly complex processes that will require decades for researchers to fully understand. In addition to the relevance of the topic for basic science, the functioning of memory has applications for the efficient use of information in designing artifical neural networks for deep learning.

One form of information storage that has attracted attention in the neuroscience literature is olfactory memory in \emph{Drosophila}, also known as the \emph{olfactory cascade}. Olfaction is the primary sense used by most insects, and the olfactory cascade has a manageable level of complexity, making it amenable to system-level research. Consequently, researchers have a general understanding of the relevant architecture and the processes through which olfactory sensations in the \emph{Drosophila} antenna communicate with the mushroom body memory center and the ways in which learned associations vary with experience.

The primary aim of this study is descriptive. Using public data that have only recently become available, this paper illustrates and discusses patterns in the actual distributions of neurons and connections in the olfactory cascade. The summary statistics and relationships identified in this article are used to derive general principles about how neurons and synapses are organized in the \emph{Drosophila}'s olfactory system. As memory is a current area of interest in the field of deep learning, organizational features of the olfactory cascade---some previously known and some illustrated in this study---are then discussed in the context of how they might apply when incorporating memory into the design of artificial neural networks.

Section \ref{Olfactory Data} provides a brief summary of the data sources used in this study. Section \ref{Olfactory Anatomy} presents a general overview of the fruit fly's processing of olfactory information based upon evidence from the Neuprint hemibrain data, a newly available dataset on the \emph{Drosophila} brain, and it relates the key features of this cognitive system to common systems of artificial intelligence. The Neuprint dataset contains a relatively complete mapping of the neurons and connections between them for half of the \emph{Drosophila} inner brain \citep{Scheffer2020}. The current study summarizes some key patterns from those data. The general descriptive features of the Neuprint data discussed in Section \ref{Olfactory Anatomy} include the numbers of neurons of different types in different regions that make up the olfactory processing system in the \emph{Drosophila} brain and the numbers of synaptic connections at different stages in the process. This descriptive analysis highlights the key role of Projection Neurons (PNs) in providing a bridge between the relatively unprocessed olfactory data coming from the Olfactory Neurons (ORNs) and the processing centers deeper in the \emph{Drosophila} brain. The analysis of the synapse counts illustrates the considerable variety across neurons in the roles they play in the processing of olfactory information. The following characteristics are highlighted:
\begin{enumerate}
    \item The networks involved in coding are relatively shallow
    \item Neurons in the same processing layer vary widely in their degrees of connectedness
    \item A large amount of communication is among neurons in the same layer
\end{enumerate}

Section \ref{Olfactory Coding and Memory} builds upon this analysis together with evidence from the experimental literature to discuss ways in which memories are stored and represented in the \emph{Drosophila} brain, emphasizing a fourth point:
\begin{enumerate}
    \setcounter{enumi}{3}
    \item The structure of data representations in storage are largely stable, with learning applying positive and negative associations to existing categories.
\end{enumerate}
The means by which learned associations are coded in this biological system are contrasted with approaches used in the deep learning literature. A sketch of a possible coding system is laid out that draws inspiration from the mapping of stimuli to responses in the memory cells of the fruit fly brain. A key element of the biological system is proposed as a design principle: the system has a generally stable memory structure and grouping of stimuli into categories. The learning process consists of mapping those categories of stimuli into conditioned responses based upon experience. Section \ref{Olfactory Conclusion} concludes.

\section{Data and Literature Description} \label{Olfactory Data}

Table \ref{table:datasets} describes the two Neuprint datasets used in this descriptive analysis, including the key fields and the numbers of cases in each. These datasets constitute the most comprehensive characterization to date of the neurons in the \emph{Drosophila} brain and the connections between those neurons. The data contain the full set of neurons and a best estimate of the numbers of connections between these neurons for the right half of the inner portion of the brain of a single female \emph{Drosophila} together with some portions outside that half inner brain. A full half of the mushroom body is included in the data, with a connectome including synapse links among neurons in that half. Neural connections that are asymmetric or that depend integrally upon coordination between left and right are not included in the data.

\begin{figure}[ht]
\centering
\includegraphics[width=0.5\textwidth,keepaspectratio]{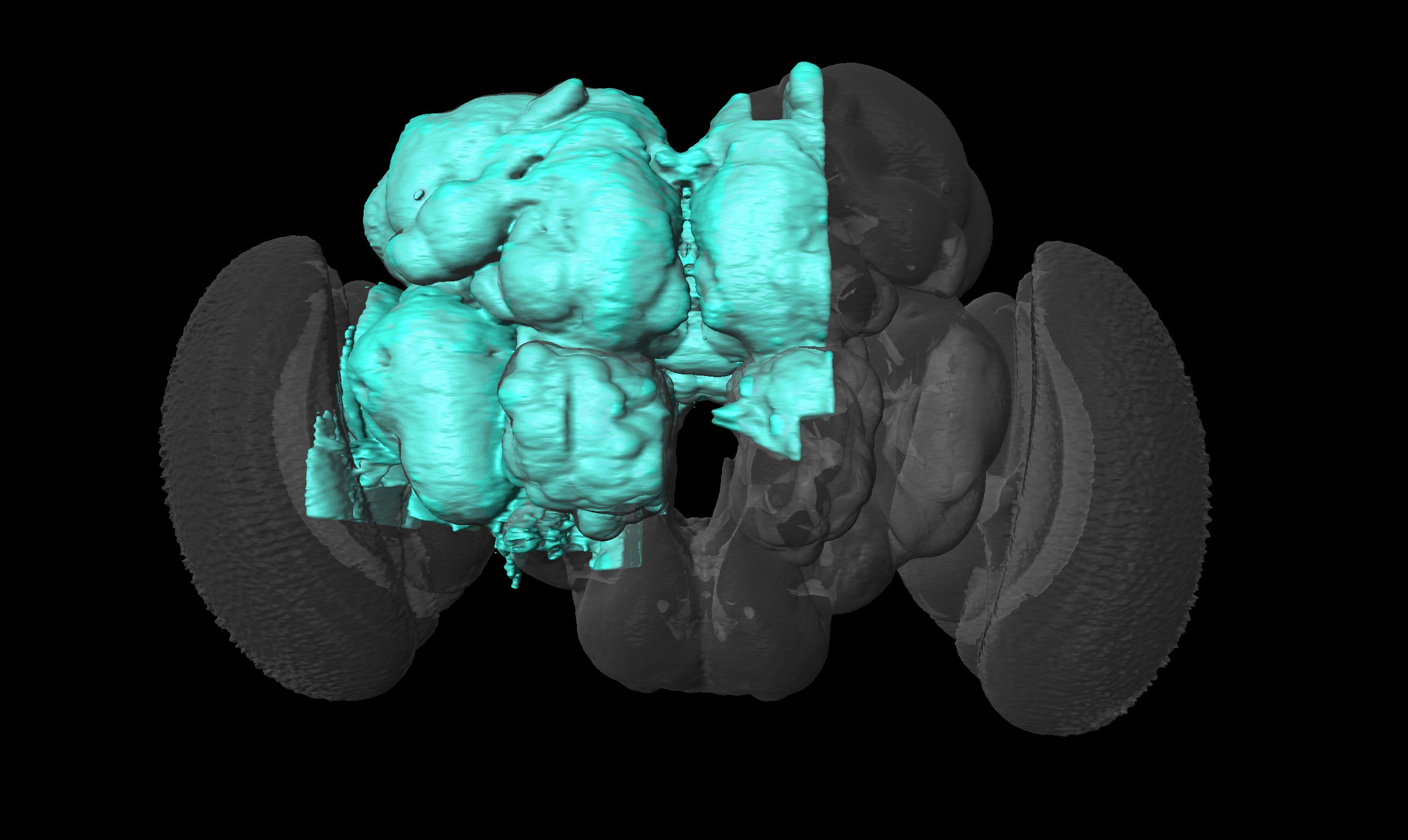} 
\begin{minipage}{\linewidth}
\flushleft \tiny Notes: The blue region illustrations the portion of the brain covered by the dataset. Image reproduced with permission from \cite{Scheffer2020}. Additional details in the text. 
\end{minipage}
\vspace{0.10 in}
\caption{Illustration of the Portion of the Brain Covered in the Neuprint Hemibrain Data}
\label{fig:hemibrain}
\end{figure}

The portion of the brain that is covered by this dataset is illustrated by the blue-shaded region in the image of the fruit fly brain in Figure \ref{fig:hemibrain}. The round shapes on the outer right and left are visual centers proximate to the \emph{Drosophila}'s eyes. The maxillary palp in the lower center and the antennae (not shown) between the palp and the visual centers contain olfactory sensors. The olfactory neurons are concentrated in the antennal lobes to the left and right of the dark body in the center of the image. The mushroom body resides above the olfactory neurons in the upper central portion of the image, and the lateral horns are located nearby, also in the upper central portion but further to the outside.

In the first dataset, the level of observation is the neuron. The region and neuropil variables include lists for each observation, describing all the regions and neuropils in which the neuron is located. The neuron name includes information about the type of neuron, with categories including ORN (Olfactory), PN (Projection), KC (Kenyon Cell), MBON (Mushroom Body Output), and DAN (Dopaminergic). The soma location is of limited usefulness for the current exercise, because the soma is often located far from the dentrites' and axons' regions of activity. In the second dataset, the level of observation is the neuron pair, describing the full set of connections leading out of and into the neurons described in the first dataset. The Start \& End IDs described in the data use the same format as the Neurons dataset but include non-neuron bodies as well as neurons not included in that dataset. Each connection has associated with it a \emph{weight} and \emph{high probability weight}, which describe the numbers of possible synapses and high probability $(p=0.85)$ synapses in that connection.

\begin{table}[ht!]
\caption{Neuprint Datasets} 
\label{table:datasets} 
\begin{threeparttable}
\centering 
\resizebox{0.5\textwidth}{!}{
\begin{tiny}
\begin{tabular}{c c c} 
\hline\hline 
Dataset & Fields & Cases \\
\hline 
\multirow{2}{*}{Neurons} & Body ID, Name, Regions \& & \multirow{ 2}{*}{24,620} \\
& Neuropils, Soma Size \& Location & \\
\\
\multirow{2}{*}{Connections} & Start \& End ID, Weight, High & 44,301,952 total, 3,583,381 with both \\
& Probability Weight & Start \& End IDs in Neuron Data \\
\hline 
\end{tabular}
\end{tiny}}
\begin{tablenotes}
\begin{subtable}{1.2\linewidth}
\vspace{0.10 in}
\tiny
\item Notes: The two datasets described here are ``Neuprint\_Neurons\_52a133.csv'' and ``Neuprint\_Neuron\_Connections\_52a133.csv.'' The first of these datasets is restricted to the subset of neurons whose status label is ``Traced'' or ``Roughly Traced'' or whose instance (name) value is not blank. The source datafiles are both components of the Neuprint hemibrain\_v1.0.1\_neo4j\_inputs database, which is described in detail in \cite{Scheffer2020}. In addition to these two sources, the current analyses uses the Region of Interest (ROI) designations from ``Neuprint\_ROIs\_52a133.csv.'' The database, which covers more than half of the \emph{Drosophila} inner brain as well as some outer components, is publicly available at \url{gs://hemibrain-release/neuprint/hemibrain_v1.0.1_neo4j_inputs.zip}. The level of observation in the first dataset is a single neuron. Name, Regions, and Neuropils are descriptive and align with common nomenclature in the literature; the fields in the dataset are labeled as uname (with a shorter variant labeled name), subregions, and neuropils. If a neuron's activity spans more than one subregion or neuropil, then multiple such locations are listed in the data. The level of observation in the second dataset is a neuron pair, with Start ID (:START\_ID(Body-ID)) and End ID (:END\_ID(Body-ID)) whose values correspond to the Body ID variable in the Neurons dataset (bodyID). Weight and High Probability Weight are integer-valued numbers of possible synapses and high probability $(p=0.85)$ synapses that were identified between those two neurons, with the signal flow from Start ID to End ID. The full connections dataset includes synapses to non-neurons as well as neurons not included in the hemibrain.
\end{subtable}
\end{tablenotes}
\end{threeparttable}
\end{table}

Beyond these two datasets, a vital source of information about the functioning of \emph{Drosophila} memory is the empirical literature. One important area of research into understanding \emph{Drosophila} olfaction and memory has been the use of experiments involving a learned odor discrimination task. A previously neutral odor is repeatedly presented to flies in conjunction with a specific reward or penalty. The odors as well as the types of rewards and penalties are varied across populations of flies and over time in order to identify specific forms of learning and unlearning. For example, one population of flies might be taught to associate a given odor with a reward, while another population---or the same population at a different time---is taught to associate it with a penalty. The experiments are performed on flies with different types of genetic modifications that inhibit the expression of certain neural pathways, thus enabling the researcher to identify the importance and roles of specific channels of signal flow in the learning process. Some key observations from this literature are described in Section \ref{Olfactory Coding and Memory}.

\section{Anatomy of the Olfactory Cascade} \label{Olfactory Anatomy}

This next section summarizes some key features of the neurons and synapses associated with the \emph{Drosophila}'s olfactory system based upon descriptive statistics from the Neuprint datasets described in detail in \cite{Scheffer2020}. Special attention is paid to differences between this biological system and common artificial neural networks. Table \ref{tab:features} summarizes the key structural features of these two forms of information processing; further discussion of the functional differences appears in Section \ref{Olfactory Coding and Memory} and Table \ref{tab:coding}. The remainder of this section first reviews their general structure and then provides evidence from the hemibrain data on each of features described in the table.

\begin{table}[ht!]
\renewcommand\thetable{2}
\caption{Characteristics of the Olfactory Cascade and Artificial Neural Networks}
\label{tab:features}
\begin{threeparttable}
\renewcommand{\arraystretch}{2}
\centering
\resizebox{0.5\textwidth}{!}{
\begin{tiny}
\begin{tabular}{p{0.075\textwidth}|p{0.2\textwidth}|p{0.20\textwidth}}
Feature & Biological System & Artificial Systems \\ \hline
Individual Neurons & Each neuron conveys continuous information about input intensity through continuous-time variation in firing rates. Neurons vary in their levels of sensitivity to different upstream nodes and in their levels of connectedness to neurons in other areas. & Generally uniform nodes that convey effectively binary information through one-shot activation. Neurons vary in their levels of sensitivity to different upstream input nodes. \\ \hline
Configuration & Separate components for initial input processing, motor responses, and memory storage. & Convolutional network layer for initial processing of input data. Later processing performed by a series of similarly-designed layers of nodes. \\ \hline
Connectedness \& Feedback & Complex network of forward and backward connections of different strengths with general tendencies of information flow. & Artificial networks are typically fully connected in the forward direction with connections occasionally skipping layers or feeding information recursively. Connections vary in strength as determined through training. \\ \hline
Network Depth & Relatively short paths from stimulus to response, with some variation across stimuli. & Generally long paths from input to classification, often with hundreds or thousands of layers of information processing; typically common depth for all inputs. \\ \hline
\end{tabular}
\end{tiny}}
\begin{tablenotes}
\begin{subtable}{0.9\linewidth}
\vspace{0.10 in}
\tiny
\item Notes: Features of individual biological neuron described in \cite{Kandel2012}. Configuration of \emph{Drosophila} memory configuration described in \cite{Aso2014a, Aso2014b, Aso2016, Boto2020, Carraher2015, Chiang2011, Li2015, Mao2009, Rains2009, Riabinina2016, Scheffer2020, Syed2015, Vosshall2000}. Levels of connectedness and feedback for \emph{Drosophila} discussed in those studies and also in the descriptive analyses presented in this paper. The shallowness of the olfactory processing system is a key finding illustrated in subsection \ref{Depth} of this study. General structure and common features of artificial neural networks are presented in \cite{GoodfellowBengioCourville2017}. On the depth of artificial neural networks, \cite{Xiao2018} note that CNNs often have hundreds or thousands of layers, and in their textbook on Deep Learning, \cite{GoodfellowBengioCourville2017} mention that, ``Empirically, greater depth does seem to result in better generalization for a wide variety of tasks,'' pg. 195. Further discussion of the functional differences appears in Section \ref{Olfactory Coding and Memory} and Table \ref{tab:coding}.
\vspace{0.10 in}
\end{subtable}
\end{tablenotes}
\end{threeparttable}
\end{table}

\subsection{Structural Overview} \label{Olfactory Structure}

\begin{figure}[ht]
\centering
    \begin{subfigure}[ht]{0.28\textwidth}
        \centering
        \includegraphics[width=\textwidth,keepaspectratio]{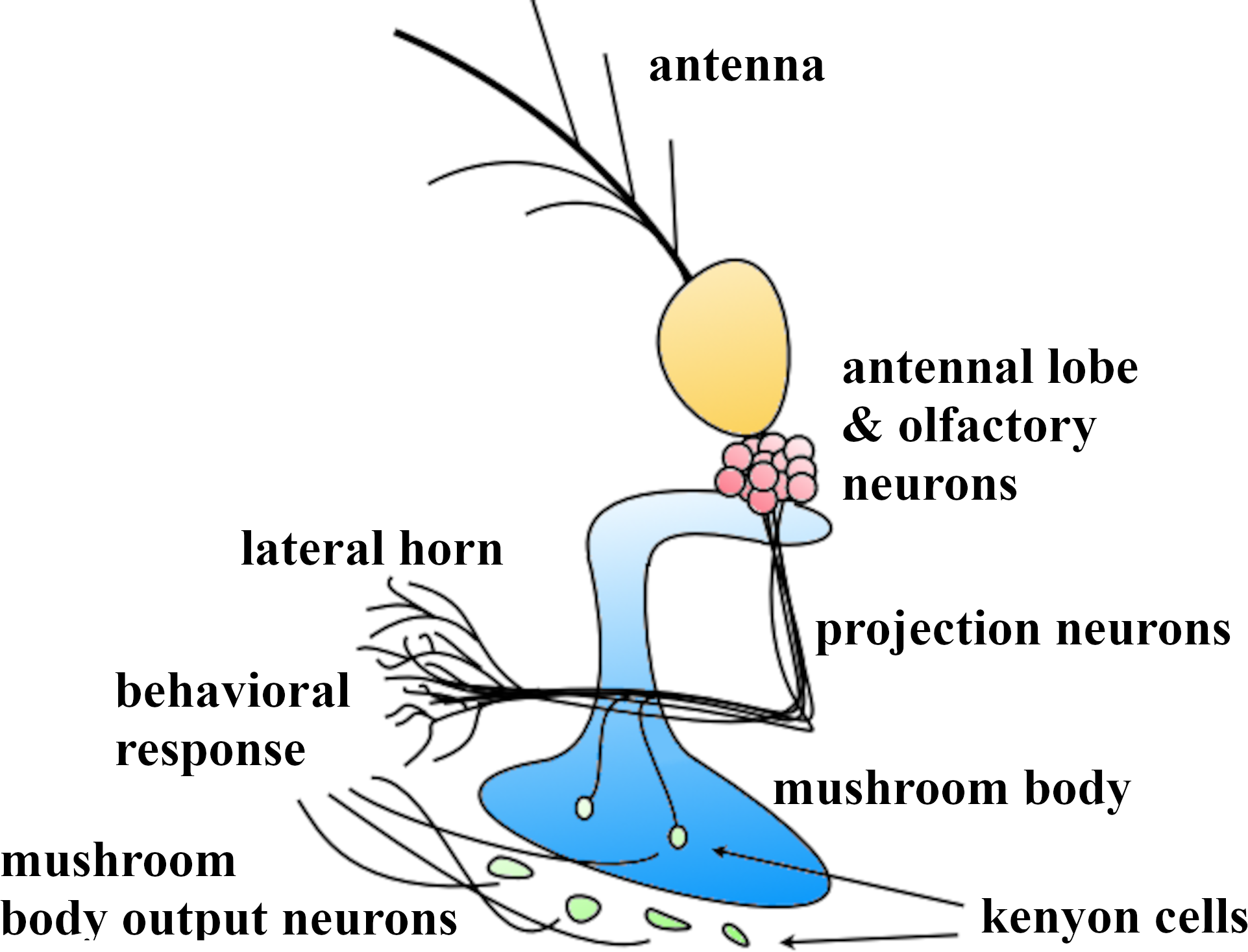} 
        \caption{The olfactory cascade} \label{fig:olfactory_cascade}
    \end{subfigure}
    \hfill
    \begin{subfigure}[ht]{0.17\textwidth}
        \centering
        \includegraphics[width=\textwidth,keepaspectratio]{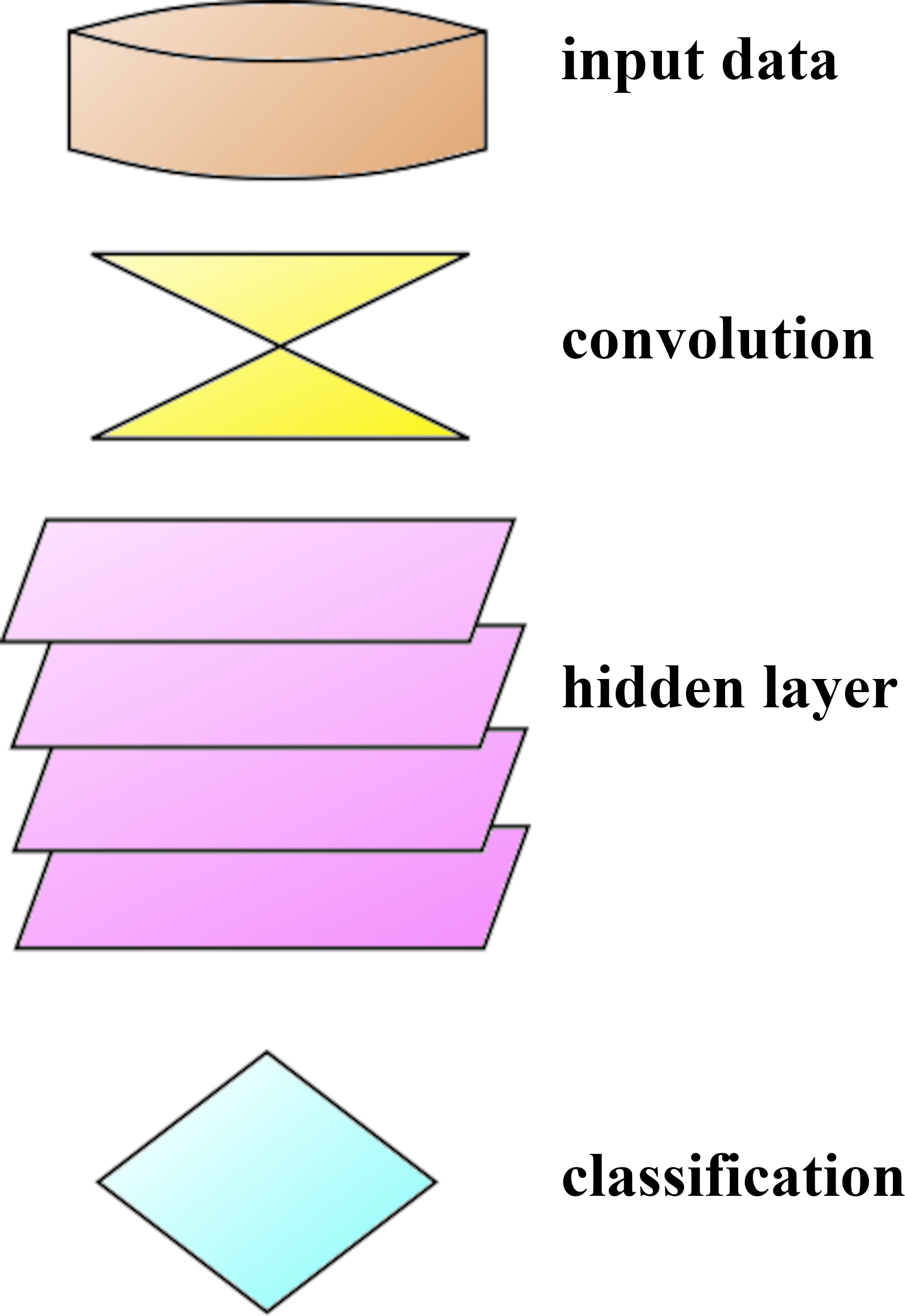} 
        \caption{Artificial neural network} \label{fig:olfactory_cnn}
    \end{subfigure}
\begin{minipage}{\linewidth}
\flushleft \tiny Notes: Figure \ref{fig:olfactory_cascade} illustrates the key anatomical components of olfactory sensation and memory in \emph{Drosophila}. Odorants are collected in the antenna. Olfactory Neurons in the antennal lobe have firing rates that depend upon the concentrations of different odorants. Projection Neurons receive that initial sensory input and transmit it to the lateral horn for innate responses and to Kenyon Cells in the mushroom body for learning and learned responses. The Kenyon Cells transmit signals to Mushroom Body Output Neurons, which pass information further downstream to motor responses. Figure \ref{fig:olfactory_cnn} illustrates the general structure of a typical feed-forward artificial neural network with a convolution step. The goal of the system is to replicate some complex set of rules that groups input cases into categories---for instance, determining what object is represented in an image based upon an input grid of pixel intensities for that image. Data received in an initial input step are transmitted to a convolutional layer. The local averaging of the convolution serves to consolidate local patterns of information and to summarize patterns in the input feed. Next, the information from the convolution layer is sent to hidden layers that perform a series of weighting, consolidation, and nonlinear threshold-based transformations to the input feed. Learning makes use of a training set of data in which the input characteristics and the final categorizations are known for each case. The weights for the connections across nodes in these hidden layers are calibrated over a training process to minimize the classification error for this training dataset. After the training is complete, the system is deployed to classify cases whose categories are not known in advance. Illustrations by author building on images and concepts found in \cite{Aso2014a, Aso2014b, Aso2016, Boto2020, Carraher2015, Chiang2011, Li2015, Mao2009, Rains2009, Riabinina2016, Scheffer2020, Syed2015, Vosshall2000} for Figure \ref{fig:olfactory_cascade} and in \cite{GoodfellowBengioCourville2017} for Figure \ref{fig:olfactory_cnn}.
\end{minipage}
\vspace{0.10 in}
\caption{Stylized illustration of signal flow in biological and artificial systems}
\label{fig:signal_flow}
\end{figure}

Figure \ref{fig:signal_flow} illustrates the stages of information processing in two systems. The orientation is flipped vertically from the hemibrain image in Figure \ref{fig:hemibrain}, the locations associated with the bottom parts of the diagram shown here appear in the upper central portion of that image of the brain. The sequence of events and linkages influencing olfactory sensation and memory in the fruit fly is known as the olfactory cascade and is shown in Figure \ref{fig:olfactory_cascade}. The process begins with odorant molecules making contact with hair-like sensilla in the antenna, shown at the top of the diagram. These molecules pass into liquid secretions out of the inward-facing parts of the sensilla. These secretions are conveyed by proteins to the dendrites of olfactory receptor neurons (ORNs). Some ORNs fire outside the brain into the suboesophageal ganglion (not shown), where they influence direct responses such as head movement and salivation. A greater portion of the ORNs fire into the antennal lobe. Neurons within the antennal lobe modulate the signal received from the ORNs. The raw and processed olfactory signals affect firing rates in projection neurons (PNs) that have dendrites in the antennal lobe and axons that terminate in the lateral horn and the mushroom body. The lateral horn transforms the data it receives into motor response signals through both innate and learned relationships. The learned reactions are developed through coordination with Kenyon Cells (KCs) within the mushroom body, which is the \emph{Drosophila}'s memory center and a key location for prioritization and related analyses. Mushroom Body Output Neurons (MBONs) retrieve information from the KCs and translate it into learned responses, so that behavior reflects innate reactions from the more direct sensori-motor channels as well as learned reactions conveyed from the MBONs. Beyond the cascade illustrated in the graph, Dopaniergic Neurons (DANs) are involved in the retrieval and modulation of information from the KCs. The number and strength of synaptic connections throughout the sensori-motor system vary in response to usage and feedback---to reinforce conditioned associations and responses that are used frequently or are associated with dopaminergic rewards or penalties. This component of the learning process is particularly important at the initial output stage to the MBONs, and the DANs play a key role in the conditioning process, incorporating feedback and modulating the responses \citep{Aso2014a, Aso2014b, Aso2016, Boto2020, Carraher2015, Chiang2011, Li2015, Mao2009, Rains2009, Riabinina2016, Scheffer2020, Syed2015, Vosshall2000}.

Figure \ref{fig:olfactory_cnn} provides a stylized representation of the corresponding input-output flow for an artificial neural network with a convolutional layer. The system receives unstructured data such as pixel intensities in an image and returns a ``classification''---\emph{e.g.,} the category of object appearing in that image. An initial convolution stage produces local weighted averages. This averaging is intended to capture general patterns in the data and relationships among pixels that are close to each other so that the output signals provide more meaningful summaries than are provided by individual pixel values. Further consolidation steps---identifying additional relationships and increasing or reducing the dimensionality of the data---are performed in series in the hidden layers. Each node in each hidden layer aggregates the input signals it receives with a weighted sum and then returns a transformation of that input by applying a nonlinear ``activation function.'' The final output layer returns a classification for the case being analyzed. Development and implementation of the system typically occur as distinct steps on separate datasets. In the development stage, alternate combinations of weights are considered for the aggregation by the hidden layers and the final output layer---using an optimization procedure to minimize the forecasting error on an initial subset of the data. For broader modeling decisions---including the assumed values for ``hyperparameters''---the performance of competing fully-trained models is compared on a separate validation sample. While some hyperparameters influence the design of the convolution stage, that early layer does not typically have weights that are determined in the training stage. A key advantage of the convolution is that it performs data consolidation tasks that are common across multiple use cases and do not require relearning for each application \citep{GoodfellowBengioCourville2017, Lecun1989}.

\subsection{Individual Neurons} \label{Olfactory Individual Neurons}

An important distinction between biological and artificial networks is in the amount of information conveyed by a single neuron. Both biological and artificial neurons vary in their levels of sensitivity to different segments in the input domain through different rates of connectively to upstream neurons. In the biological case, this variation arises through different numbers and strengths of synapses connecting the neurons, while in the artificial case, it is implemented through different weights of connections that are calibrated in the training stage. Given the upstream signals that it aggregates, each biological neuron conveys continuous information about input intensity through continuous-time variation in firing rates. By contrast, artificial networks generally consist of uniform nodes that convey essentially binary information through one-shot activation \citep{GoodfellowBengioCourville2017, Luo2018, Reardon2018, Sinz2019, Zador2019}. Much of the complexity of artificial neural networks is contained in the many layers of transformations---often hundreds or thousands---involved in translating inputs into learned classifications or relationships \citep{GoodfellowBengioCourville2017, Xiao2018}.

Spiking neural networks (SNNs) constitute one approach that researchers have explored to enable artificial neurons to produce continuously varying output \citep{Deger2014, Maass1997, Martin2021, Muscinelli2019, Panda2020, Tavanaei2017, Wang2014}. While SNNs achieve this goal, they have been found to be slow and resource-intensive. Outside of spiking systems that incorporate this costly time element, researchers have not yet found success with neurons that perform univariate or multivariate nonlinear transformations to produce continuously varying output---the most common and successful artificial neural networks use activation functions that effectively discretize signals.

An additional neuron-level distinction between the biological and artificial systems is in the presence of inhibitory signals in biological systems. While negative weights in principle enable artificial systems to transmit inhibitory signals, biological systems make more extensive use of this mechanism. Some researchers have explored explicitly modeling this process of inhibition in order to promote sparseness and to make individual neurons more expressive \citep{Cao2018, Isaacson2011, King2013, Seung2018b, Seung2018a}.

\subsection{Configuration} \label{Olfactory Configuration}

Table \ref{table:location} describes the distribution of neurons in the neuron-level Neuprint dataset by type across locations in the \emph{Drosophila} brain. Each row shows the number of neurons of that type (ORN, PN, KC, MBON, Dopaminergic, and Other) that are associated with each of the different region and neuropil locations listed. These regional and neuropil associations are overlapping, so that the sums of the other columns exceed the totals. There are two levels of association with the Antennal Lobe---the region and the neuropil, where all neurons associated with the Antennal Lobe region are a subset of those associated with the neuropil. Lateral Horn associations are only with a neuropil and none with a region, and Mushroom Body associations are only with the region and none with the neuropil.

As the first row of the table illustrates, all Olfactory Neurons are located within the Antennal Lobe Neuropil, and none are associated with the Lateral Horn or Mushroom Body. All Kenyon Cells are located within the Mushroom Body, and none are associated with the Antennal Lobe or Lateral Horn. The role of the Projection Neurons in the overall structure of the olfactory cascade is demonstrated by their appearance in each of these different locations. Nearly all (97.4\%) of the Mushroom Body Output Neurons are associated with the Mushroom Body; of those, 13 (16.7\%) are also associated with the Lateral Horn, and 3 (3.8\%) are also associated with the Antennal Neuropil. Of the 339 Dopaminergic Neurons in the hemibrain data, 322 (95.0\%) are associated with the Mushroom Body, and 6 (1.8\%) are associated with the Lateral Horn.

\begin{table}[ht]
\renewcommand\thetable{3}
\caption{Neurons by Type and Location} 
\begin{threeparttable}
\centering 
\resizebox{0.5\textwidth}{!}{
\begin{tabular}{c c c c c c} 
\hline\hline 
\multirow{ 2}{*}{Neuron Type} & \multirow{ 2}{*}{Total} & \multicolumn{2}{c}{Antennal Lobe} & Lateral Horn & Mushroom \\
& & Region & Neuropil ex Region & Neuropil & Body Region \\
\hline 
Olfactory & 1,474 & 60 & 1,414 & 0 & 0 \\
Projection & 524 & 68 & 345 & 295 & 245 \\
Kenyon & 2,033 & 0 & 0 & 0 & 2,033 \\
Mushroom Body Output & 78 & 0 & 3 & 13 & 76 \\
Dopaminergic & 339 & 0 & 0 & 6 & 322 \\
Other & 20,172 & 138 & 345 & 2,666 & 2,225 \\
\hline 
\end{tabular}}
\begin{tablenotes}
\begin{subtable}{.65\linewidth}
\vspace{0.10 in}
\tiny
\item Notes: Author's tabulation from Neuprint Neurons file. The values presented in this table are total numbers of neurons of each type in the hemibrain data (in the Total column) and numbers associated with different regions of the fly brain (in the later columns). Neuron types are distinct and are identified based upon neuron name. Olfactory, Projection, Kenyon, and Mushroom Body Output neuron types are those whose names include ``ORN,'' ``PN,'' ``KC,'' or ``MBON,'' respectively, and Dopaminergic neuron type is one whose name includes one of ``PP,'' ``PAM,'' or ``DAN.'' These categorizations are largely consistent with those used in \cite{Scheffer2020}; they are mutually exclusive, and ``Other'' consists of neurons not falling into these categories. Each neuron's regions and neuropils of interest were cross-referenced against the Neuprint Region of Interest (ROI) file, and groupings of subregions into larger categories were determined based upon lists appearing in \cite{Scheffer2020}. Antennal lobe consists of AL-DC(R), AL(L), and AL(R) subregions. A neuron with one of those subregions included in its ``regions'' list is classified as being associated with the Antennal Lobe region, and a neuron with one of those subregions included in its ``neuropils'' list is classified as being associated with the Antennal Lobe neuropils. The Lateral Horn subregion included is LH(R). Subregions associated with the Mushroom Body include a'1(R), a1(R), a'2(R), a2(R), a'3(R), a3(R), a'L(L), aL(L), a'L(R), aL(R), b'1(R), b1(R), b'2(R), b2(R), b'L(L), bL(L), b'L(R), bL(R), CA(L), CA(R), dACA(R), g1(R), g2(R), g3(R), g4(R), g5(R), gL(L), gL(R), lACA(R), PED(R), and vACA(R). These region and neuropil associations are not mutually exclusive. Matches were identified as those containing these neuron type or subregion strings, and the resulting lists and associations were checked against those appearing in \cite{Scheffer2020} and on the Janelia hemibrain site.
\vspace{0.10 in}
\end{subtable}
\end{tablenotes}
\end{threeparttable}
\label{table:location} 
\end{table}

\begin{figure}[ht]
    \centering
    \begin{subfigure}[ht]{0.22\textwidth}
        \centering
        \includegraphics[width=\textwidth,keepaspectratio]{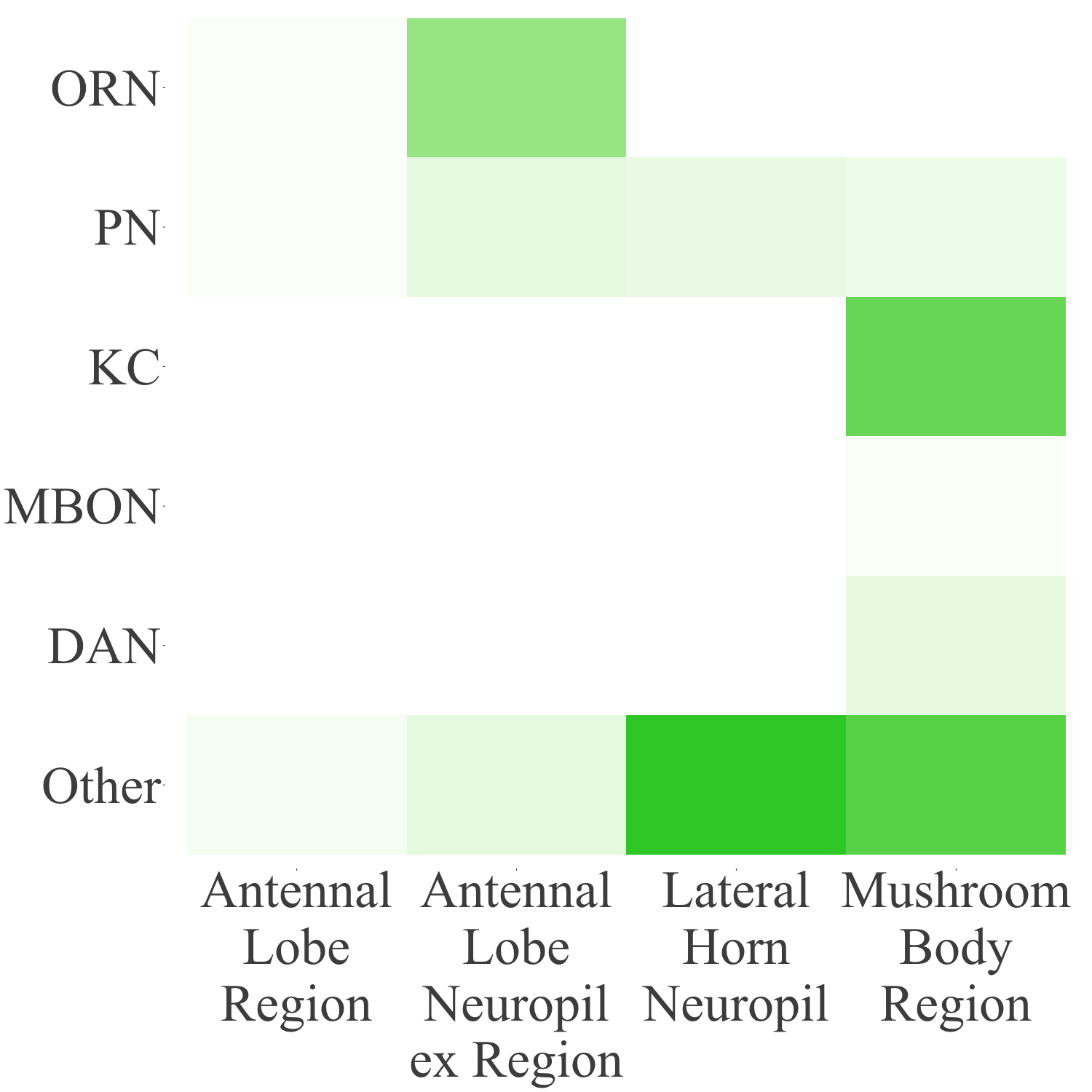} 
        \caption{Neurons by Type and Location} \label{fig:location}
    \end{subfigure}
    \begin{subfigure}[ht]{0.22\textwidth}
        \centering
        \includegraphics[width=\textwidth,keepaspectratio]{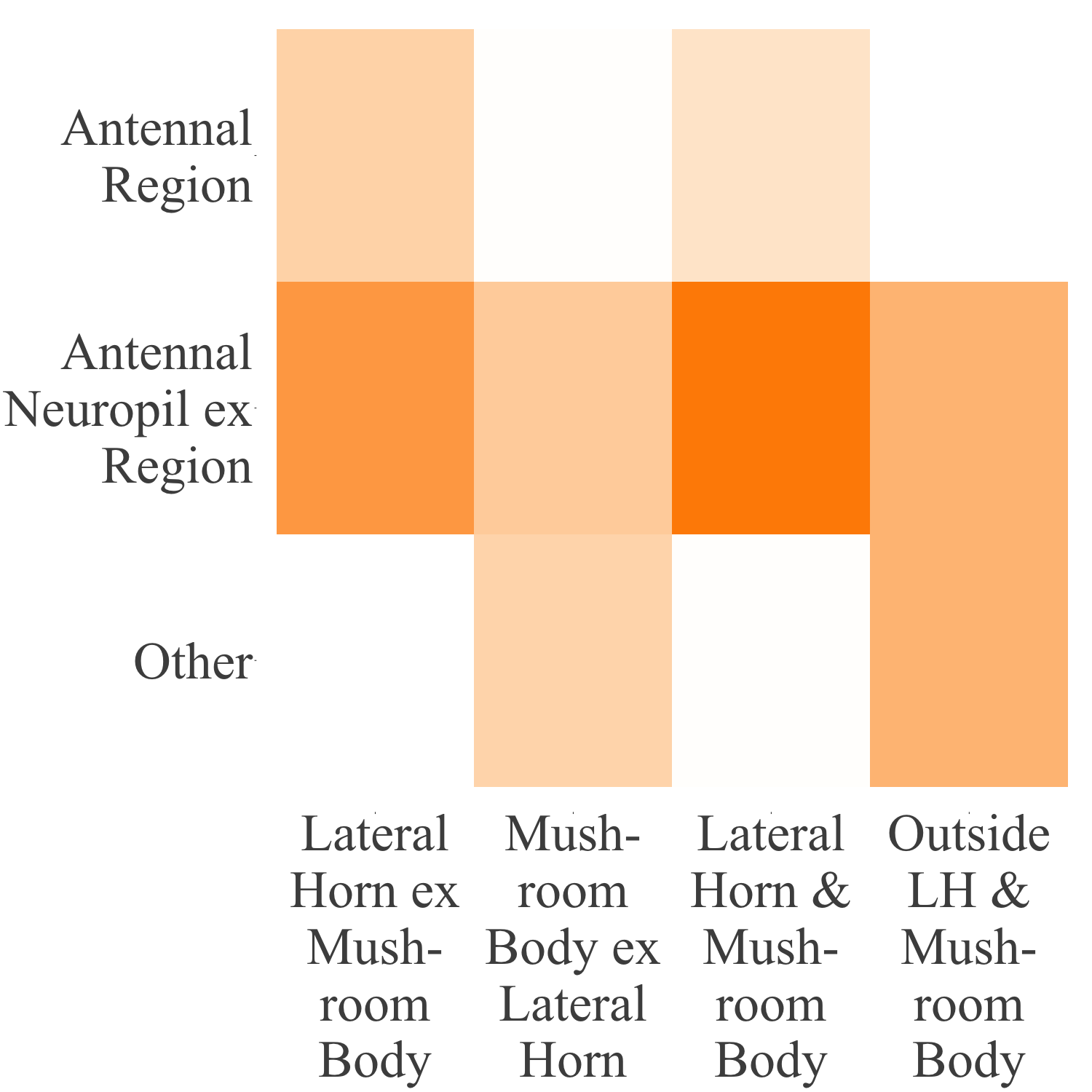} 
        \caption{Overlapping Region Classifications for Projection Neurons} \label{fig:overlapping}
    \end{subfigure}
\begin{minipage}{\linewidth}
\flushleft \tiny Notes: Values from Figure \ref{fig:location} are an illustration of data presented in Table \ref{table:location}. Values from Figure \ref{fig:overlapping} illustrate the total numbers of Projection Neurons whose region and neuropils associations fall into each of the 12 categories shown. The sample for Figure \ref{fig:overlapping} is restricted to the 524 cases shown in Table \ref{table:location} with Neuron Type of ``Projection.'' Regional classifications use the same definitions in both cases. The rows in that second graph identify three mutually exclusive categories depending upon whether the neuron was associated with an antennal subregion in the ``region'' variable, in the ``neuropil'' variable but not in the region variable, or in neither. The columns identify four mutually exclusive categories depending upon whether the neuropil variable included LH(R), the Lateral Horn subregion included in the data (yes for columns 1 and 3 and no for columns 2 and 4) and whether the region variable included one of the Mushroom Body locations (yes for columns 2 and 3 and no for columns 1 and 4). Because the regional and neuropil classifications in this table are mutually exclusive, the values in the 12 cells sum to 524, the number of Projection Neurons listed in Table \ref{table:location}.
\end{minipage}
\vspace{0.10 in}
    \caption{Illustration of Neurons by Type and Location} \label{fig:location_overlapping}
\end{figure}

The values presented in Table \ref{table:location} are illustrated graphically in Figure \ref{fig:location}, with darker shading corresponding to higher values. The rows and columns of that graph are the same as in the table but with the ``Total'' category omitted. Figure \ref{fig:overlapping} delves further into the regional breakdowns of these Projection Neurons that provide bridges between the Antennal Lobe and the Mushroom Body by illustrating the ways in which their region and neuropil associations overlap. The three rows split the set of PNs into those that fall within the Antennal Lobe region, other parts of the neuropil, and outside of the Antennal Lobe neuropil. The columns show the numbers of these Projection Neurons that are also associated with the Lateral Horn, the Mushroom Body, both, or neither. Unlike in the previous table and figure, mutually exclusive categories are shown here, so that each of the $524$ PNs is assigned to exactly one of the $12$ cells. For ease of exposition, in the tile and bar graphs shown in Figures \ref{fig:location_overlapping}, \ref{fig:spn}, \ref{fig:syntype}, and \ref{fig:pnsyn}, green shading is used for comparisons across all neuron types, and orange shading is used when only PNs are being compared.

As Table \ref{table:location} and Figure \ref{fig:location} illustrate, unlike ORNs and KCs, Projection Neurons are spread across a diversity of locations, and considerable overlap in terms of region and neuropil associations. Nevertheless, some distinct patterns can be seen. Most of the PNs that are associated with the Antennal Neuropil are also associated with the Lateral Horn. Of the 68 PNs in the Antennal Region in the first row, 67 (98.5\%) are associated with the Lateral Horn. Of the 111 PNs that are outside of the Antennal Neuropil entirely in the third row, only one (0.9\%) is associated with the Lateral Horn. Among the 345 PNs in the second row that are associated in the Antennal Neuropil outside the Antennal Region, 227 (65.8\%) are associated with the Lateral Horn. Additionally, the Lateral Horn can be seen as a key stopping point in the coding process. Most of the PNs in the Antennal Lobe Neuropil are associated with the Lateral Horn, and most of the PNs associated with the Mushroom Body are also associated with the Lateral Horn. This pattern of associations is consistent with the Lateral Horn being a chief recipient of raw olfactory information and mediating the process of coding and retrieving olfactory memory through connections with the Mushroom Body.

As noted in subsection \ref{Olfactory Structure}, the nodes in a typical artificial neural network exhibit some variation. Specifically, convolutional layers consolidate data and identify relationships in a standardized way that does not depend upon trained parameters. In that sense, the convolutional stage resembles the initial processing that is performed by ORNs in the olfactory cascade. The trained hidden layers that are further downstream in an artificial network perform processing tasks that are analogous to those performed by the PNs and KCs in the olfactory cascade; however, these hidden layers and their component nodes are not differentiated like the neurons and neuropils are in the fly's olfactory system. The \emph{Drosophila} brain's incorporation of dopaminergic feedback through DANs and the interactive formation and strengthening of synapses to MBONs and throughout the system differ from conventional supervised learning with artificial neural networks that involves sequential training, validation, and deployment---and more closely resemble online learning approaches.

\subsection{Connectedness and Feedback} \label{Connectedness}

\begin{table}[ht]
\renewcommand\thetable{4}
\caption{High Probability Outgoing and Incoming Synapses per Neuron} 
\begin{threeparttable}
\centering 
\resizebox{0.45\textwidth}{!}{
\begin{tiny}
\begin{tabular}{c c c c c} 
\multicolumn{5}{c}{Panel A: All Neurons by Type} \\
\hline\hline 
Neuron Type & Outgoing & Incoming & Outgoing per Incoming & t-statistic \\
\hline 
Olfactory & 325.2 & 98.4 & 3.3 &  49.8*** \\
Projection & 2,234.6 & 852.5 & 2.6 & 15.4*** \\
Kenyon & 592.6 & 454.1 & 1.3 & 89.2*** \\
Mushroom Body Output & 4,082.4 & 7,471.9 & 0.5 & -5.69*** \\
Dopaminergic & 1,080.0 & 987.0 & 1.1 & 0.89** \\
Other & 1,123.6 & 503.2 & 2.2 & 68.5*** \\
\\
\multicolumn{5}{c}{Panel B: Projection Neurons by Location} \\
\hline\hline 
Region & Outgoing & Incoming & Outgoing per Incoming & t-statistic \\
\hline 
Antennal Region & 3,098.8 & 1,160.3 & 2.7 & 9.48*** \\
Antennal Neuropil ex Region & 2,674.9 & 1,032.1 & 2.6 & 13.4*** \\
Lateral Horn & 3,437.7 & 1,206.6 & 2.8 & 16.4*** \\
Mushroom Body & 3,420.2 & 1,298.6 & 2.6 & 12.7*** \\
\hline
\end{tabular}
\end{tiny}}
\begin{tablenotes}
\begin{subtable}{1.1\linewidth}
\vspace{0.10 in}
\tiny
\item Notes: Author's computation from Neuprint Neurons and Connections files. This table shows the average numbers of outgoing and incoming high probability synapses per neuron---and the ratio of those two averages---presented separately for different subsets of the neurons in the hemibrain data. The t-statistic comes from a one-sample t-test of the null hypothesis that the Outcoming minus Incoming difference is equal to zero, assuming independent draws across the neurons. Asterisks indicate statistical significance (*** for 1\%, ** for 5\%, and * for 10\%). The sets of neurons, Neuron Type designations, and location categories in panels A and B are the same as in Table \ref{table:location}, and as in that table, the regions are not mutually exclusive. For the Outgoing column, Body ID in the Neurons dataset was matched with Start ID in the Connections data. The number of high probability synapses was summed in this merged dataset across all neurons within the specified type and divided by the number of neurons of that type. All high probability synapses are included in these averages, regardless of whether the target was a neuron in the hemibrain data. For the Incoming column, a similar calculation was performed but matching with End ID in the Connections data. The Outgoing per Incoming column is the ratio of the previous two columns. Panel A shows breakdowns by Neuron Type, and Panel B is a breakout of the second row of Panel A, with average numbers of outgoing and incoming high probability synapses per Projection Neuron, presented separately for neurons with different regional associations. TThe first row of panel B shows averages for the 68 PNs associated with the Antennal Lobe Region---the same 68 neurons shown in the first row of Figure \ref{fig:overlapping}. The second row shows averages for the 345 PNs associated with the Antennal Lobe Neuropils but not the Region, as in the second row of Figure \ref{fig:overlapping}. The third row shows averages for the 294 PNs associated with the Lateral Horn, as in the first and third columns of Figure \ref{fig:overlapping}, and the fourth row shows averages for the 245 PNs associated with the Mushroom Body, as in the second and third columns of Figure \ref{fig:overlapping}. All four sets of PNs have numbers of outgoing and incoming high probability synapses that exceed the Table \ref{table:location} averages for PNs in the data, because the categories here exclude the 70 less connected PNs that are not associated with any of these four regions (as in the lower right-hand corner of Figure \ref{fig:overlapping}).
\end{subtable}
\end{tablenotes}
\end{threeparttable}
\label{table:spn} 
\end{table}

\begin{figure}[ht]
    \centering
    \begin{subfigure}[ht]{0.22\textwidth}
        \centering
        \includegraphics[width=\textwidth,keepaspectratio]{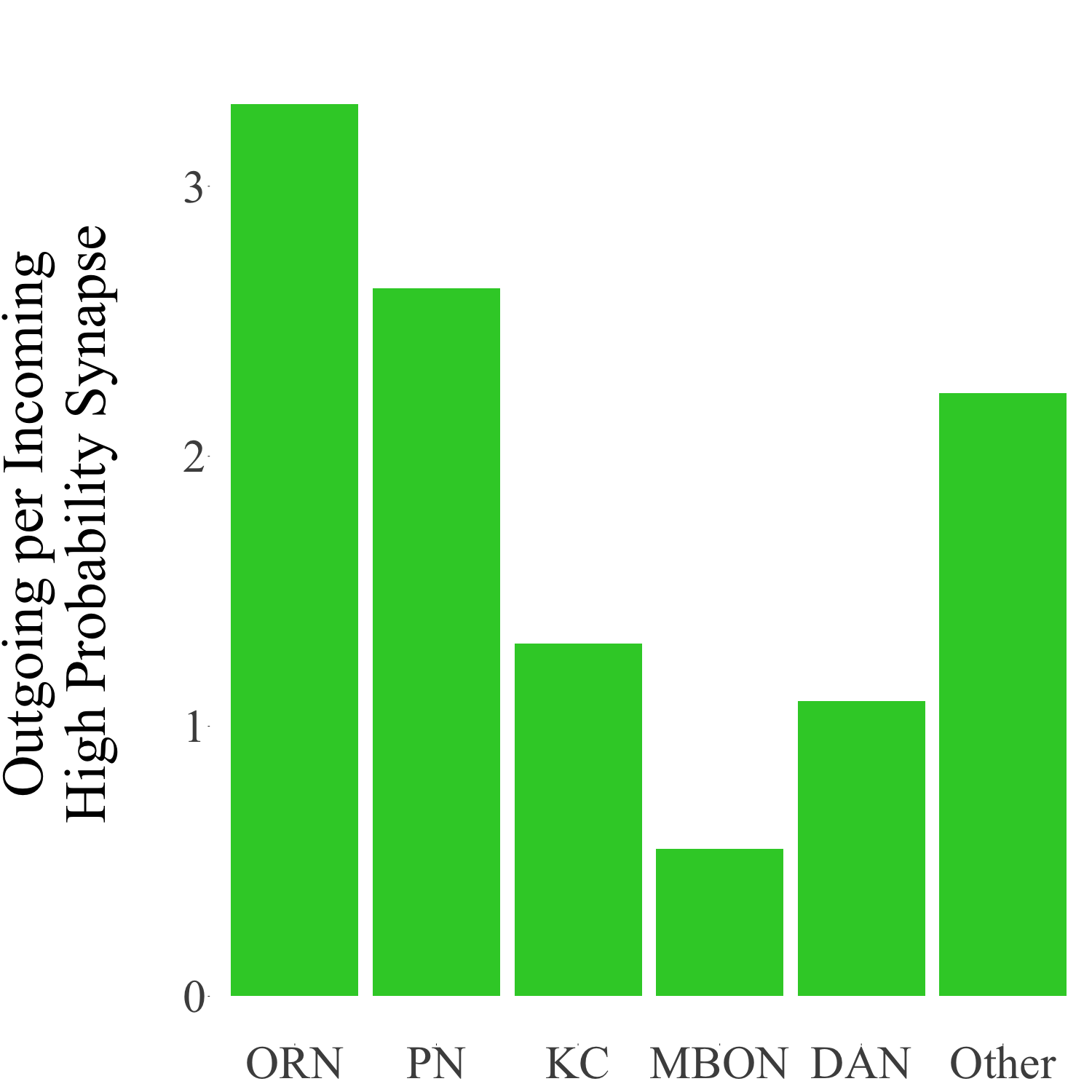} 
        \vspace{0.17 in}
        \caption{by Neuron Type} \label{fig:spntype}
    \end{subfigure}
    \hfill
    \begin{subfigure}[ht]{0.22\textwidth}
        \centering
        \vspace{0.17 in}
        \includegraphics[width=\textwidth,keepaspectratio]{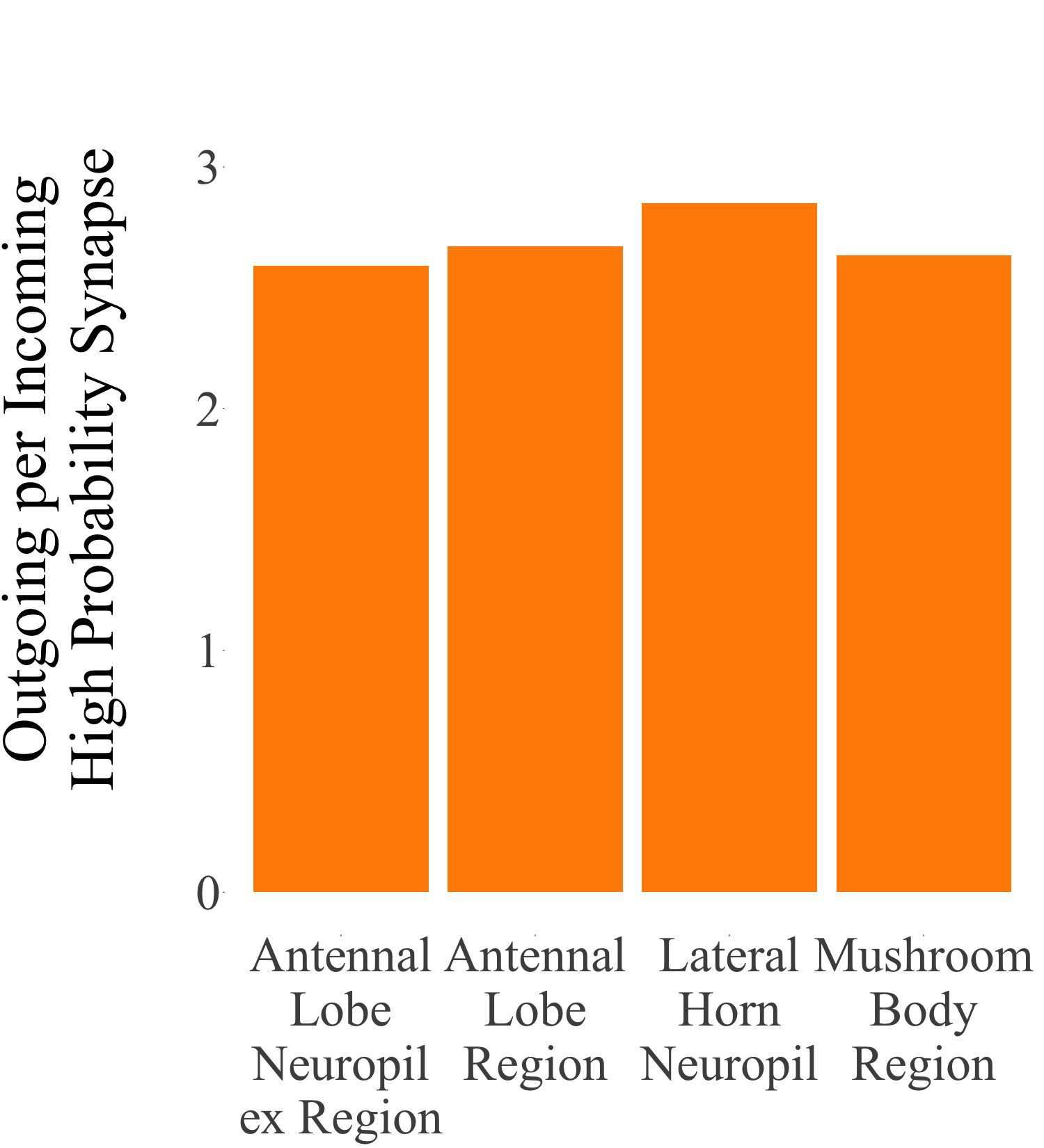} 
        \caption{Projection Neurons by Region} \label{fig:synregion}
    \end{subfigure}
\begin{minipage}{\linewidth}
\flushleft \tiny Notes: Illustration of Outgoing per Incoming column from Table \ref{table:spn}.
\end{minipage}
\vspace{0.10 in}
    \caption{Outgoing per Incoming High Probability Synapses} \label{fig:spn}
\end{figure}

The previous section's discussion about the locations of neural activity provides suggestive information about different neurons' roles in sensation and memory. Next, we examine the counts of synapses, as in the tables and figures below, to obtain a clearer picture of the importance of the roles of the different components in this system and how they interact. The first panel of Table \ref{table:spn} shows the numbers of outgoing and incoming high probability synapses per neuron for ORNs, PNs, KCs, MBONs, Dopaminergic, and Other types of neurons. The ratio of the outgoing and incoming averages is shown for each type, and the last column presents a t-statistic for a test of the null hypothesis that the difference of outgoing minus incoming has a mean of zero. The values of the ratio of Outgoing per Incoming High Probability Synapses are also presented visually in Figure \ref{fig:spntype}. The synapse counts presented here include links to neurons outside of the Neuprint data. As the counts illustrate, the number of high probability synapses per neuron is highest for Mushroom Body Output Neurons and is lowest for Olfactory Neurons. If the observations are assumed to be independent but with differing variances across neuron types, the null hypothesis of equal means for the Outgoing + Incoming sum is rejected at the 5\% level for each of the pairs of neuron types shown, with the least significant difference between DAN and Other ($t=3.08$) and the most significant difference between ORN and KC ($t=-57.1$). Additionally, the Outgoing per Incoming High Probability Synapse ratios in the right-hand column are consistent with the different neuron types' roles in the olfactory cascade. ORNs, at the sensory end of the process, have 3.3 outgoing high probability synapses per incoming high probability synapse. The ratio declines as information moves through this pipeline, with ratios of 2.6 for PNs, 1.3 for KCs, and 0.5 for MBONs. The average KC has more outgoing than incoming high probability synapses. The MBONs, which are less numerous than ORNs, PNs, or KCs, show higher levels of connectedness than those other types overall. The MBONs' high level of input relative to output high probability synapses is consistent with them consolidating information and distilling it into simpler signals that are relevant for specific behavioral responses.

PNs, KCs, MBONs, and DANs are inner neurons whose primary source of input is from other neurons. By contrast, ORNs are sensory neurons, and most of the information received by ORNs is not synaptic but consists of chemical signals from odorants collected in the antennae. The ``incoming'' numbers for ORNs do not reflect this primary source of data and solely represent synaptic connections. These incoming connections serve to modulate the ORNs' sensory input signals and to incorporate sensory data from other ORNs. For the average ORN with 98.4 incoming synapses, 73.4 (74.6\%) of those are linked to other neurons in the Neuprint hemibrain data. Of those, 47.3 (64.4\%) correspond to a variety of other neuron types outside of the ORN $\rightarrow$ PN $\rightarrow$ KC cascade. Among the incoming synapses that were identified to be part of the cascade, 24.5 (94.0\%) were from ORNs, 1.6 (6.0\%) were from PNs, and zero were from KCs, MBONs, or DANs.

To better understand the role of Projection Neurons' synapses in the olfactory cascade, the second panel of Table \ref{table:spn} along with Figure \ref{fig:synregion} summarize the numbers of outgoing and incoming high probability synapses per PN as they vary with the location of the PN. The structure is similar to that of the first panel, with the third column of values presenting the ratio of the prior two columns and the values for the ratio also shown in the figure. The outgoing and incoming totals in Panel B all exceed the corresponding averages for PNs from Panel A, because PNs located outside of these categories (as in the lower right-hand cell of Figure \ref{fig:overlapping}) are not included in panel B of the table. As the panel and figure illustrate, there is relatively little variation across regions and neuropils in the counts of outgoing or incoming high probability synapses per PN. There are generally more Outgoing and Incoming high probability synapses among PNs associated with the Lateral Horn and Mushroom Body than with the Antennal Region or Neuropil, and among the 454 PNs in the four regions presented, the difference in the average number of Outgoing plus Incoming high probability synapses between those in the latter two categories and those in the first two is $2,043.9, t=5.89^{***}.$

\begin{table*}[ht]
\renewcommand\thetable{5}
\caption{High Probability Synapses Between Neurons by Type} 
\begin{threeparttable}
\centering 
\resizebox{\textwidth}{!}{
\begin{tiny}
\begin{tabular}{c c c c c c c c | c} 
\hline\hline 
& & \multicolumn{6}{c}{Ending Neuron Type} \\
& & \textbf{Olfact-} & \textbf{Project-} & \textbf{} & \textbf{Mushroom} & \textbf{Dopamin-} &  & \\
& & \textbf{ory} & \textbf{ion} & \textbf{Kenyon} & \textbf{Body Output} & \textbf{ergic} & Other & Total \\
\hline
& \textbf{Olfactory} & \textbf{36,156} & \textbf{105,741} & \textbf{0} & \textbf{0} & \textbf{0} & 110,446 & 252,343 \\
Start- & \textbf{Projection} & \textbf{2,319} & \textbf{32,253} & \textbf{116,782} & \textbf{1,272} & \textbf{637} & 264,467 & 417,730 \\
ing & \textbf{Kenyon} & \textbf{0} & \textbf{2,551} & \textbf{503,407} & \textbf{281,682} & \textbf{130,142} & 157,721 & 1,075,503 \\
Neuron & \textbf{Mushroom Body Output} & \textbf{0} & \textbf{650} & \textbf{11,619} & \textbf{14,867} & \textbf{5,484} & 83,347 & 115,967 \\
Type & \textbf{Dopaminergic} & \textbf{0} & \textbf{196} & \textbf{111,475} & \textbf{43,358} & \textbf{12,720} & 69,008 & 236,757 \\
& Other & 69,698 & 164,839 & 119,845 & 68,767 & 78,101 & 7,053,617 & 7,554,867 \\
\hline
& Total & 108,173 & 306,230 & 863,128 & 409,946 & 227,084 & 7,738,606 & 9,653,167 \\
\hline 
\end{tabular}
\end{tiny}}
\begin{tablenotes}
\begin{subtable}{0.95\textwidth}
\vspace{0.10 in}
\tiny
\item Notes: Author's computation from Neuprint Neurons and Connections files. The values in the grid show the numbers of high probability synapses summed across all neuron-to-neuron pairs in the Connections dataset with Neuron Type for the starting neuron as specified by the row label and Neuron Type for the ending neuron as specified by the column label. The matching of neurons to connections is similar to that described in Table \ref{table:spn}, but two separate matches are performed---one for Start ID and one for End ID. Unlike in Table \ref{table:spn}, the numbers of high probability synapses shown here are not normalized by the numbers of neurons, and the totals only include high synapses in which both the source and the target are neurons in the hemibrain dataset.
\end{subtable}
\end{tablenotes}
\end{threeparttable}
\label{table:syntype} 
\end{table*}

\begin{figure}
\centering
\includegraphics[width=0.45\textwidth,keepaspectratio]{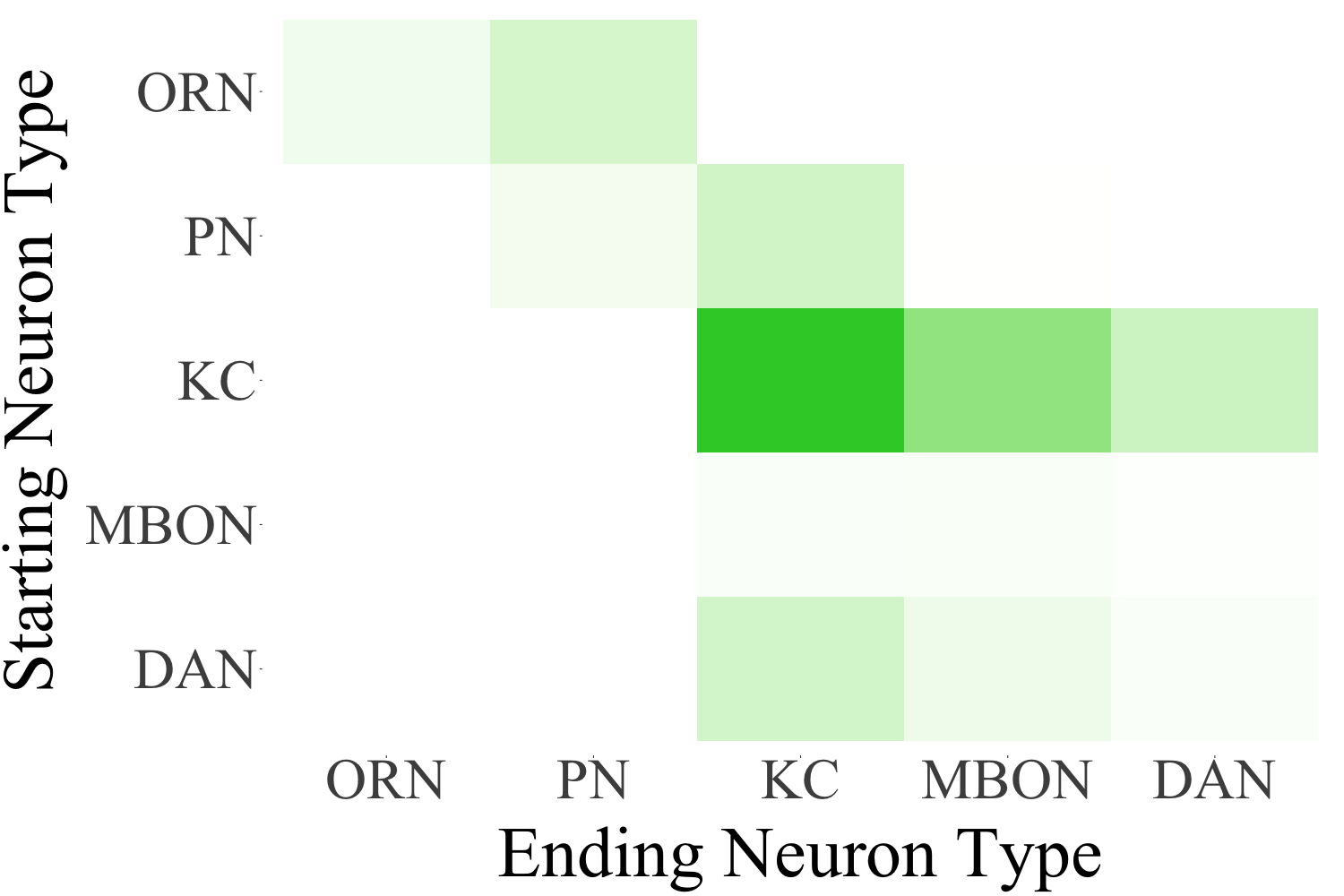} 
\begin{minipage}{\linewidth}
\flushleft \tiny Notes: Illustration of bolded values (excluding ``Other'' category) from Table \ref{table:syntype}. 
\end{minipage}
\vspace{0.10 in}
\caption{Illustration of High Probability Synapses Between Neurons by Type}
\label{fig:syntype}
\end{figure}

Next, Table \ref{table:syntype} and Figure \ref{fig:syntype} provide a two-way breakdown of which types of neuron pairs these high probability synapses tend to connect. The rows correspond to the starting neuron types with axons sending outgoing signals, and the columns corresond to ending neuron types with dendrites receiving the signals. The counts shown in the table are total numbers of high probability synapses without scaling by numbers of neurons, giving the number of high probability synapses between axons of that starting type and dendrites of that ending type. Only identified neuron-to-neuron synapses are included in these counts, and cases are excluded if the start or end of the connection is not a neuron identified in the Neuprint hemibrain data. The numbers in \textbf{bold} in the table indicate connections in which both the starting and ending neuron type is associated with key elements of the olfactory cascade discussed here (ORNs, PNs, KCs, MBONs, DANs), and those same bold values are illustrated in the figure.

As the table and figure illustrate, there are no direct connections between Olfactory Neurons and Kenyon Cells, Mushroom Body Output Neurons, or Dopaminergic Neurons in either direction. The PNs serve as a primary channel for conveying information from ORNs to KCs. The counts of high probability synapses in the different cells are consistent with this general pattern of information flow. Among the types of neurons relevant to olfactory memory, ORNs send primarily to PNs but also communicate with other ORNs. PNs send mainly to KCs but also to other PNs, with some amount of feedback to ORNs and some amount directly to MBONs and DANs. The KCs mainly send information to other KCs and onward to MBONs and DANs, with slight amounts transmitted as feedback to PNs. The MBONs primarily send their output outside the system; however, they do transmit information amongst themselves, to KCs and DANs, and in small amounts to PNs. The DANs primarily transmit their signals to KCs but also to MBONs, amongst themselves, and in small amounts to PNs. Additionally, there is a considerable amount of communication outside the set of ORNs, PNs, KCs, MBONs, and DANs (the non-bold numbers in the table). KCs are the most insulated, with only 14.7\% of their outgoing and 13.9\% of their incoming high probability synapses accounted for by types outside this set. ORNs and PNs are more outwardly oriented, with between 40\% and 65\% of their incoming and outgoing high probability synapses corresponding to connections to Other neuron types. The DANs are in between, with connections to Other neuron types accounting for 29.1\% of their outgoing and 34.4\% of their incoming high probability synapses. MBONs predictably exhibit a distinct pattern in which 71.9\% of their outgoing high probability synapses point outside these neuron types that are relevant to olfactory memory, but only 16.8\% of their incoming high probability synapses come from outside.

It is worth noting that the synapse counts do not provide an exhaustive description of the ways in which neurons in the olfactory cascade communicate. \cite{Liu2016} find that, in addition to the chemical synapses typically examined in brain research, ``gap junctions'' or electrical synapses exist between Kenyon Cells. Relative to chemical synapses, these junctions provide simple, fast, and non-discrete transmission of information, and the authors find that they play a vital role in memory encoding.

Overall, two key differences can be observed in the levels of connectedness between this olfactory system and corresponding artificial systems. First, in the biological system, feedback is common. In artificial networks, the passage of information from downstream to upstream nodes is relatively rare. In some of the most common forms of artificial networks, including the standard CNN illustrated in Figure \ref{fig:olfactory_cnn}, data pass through successive layers of the network in a sequential fashion, so that signal flow proceeds exclusively in the downstream direction.\footnote{One form of feedback that does occur with some frequency in artificial settings appears in Recurrent Neural Networks, in which nodes send output directly to themselves to combine with other data elements from the same observation \citep{GoodfellowBengioCourville2017, Lipton2015}.} Second, the configuration of biological neurons into task-specific modules interacts with connectedness. Communication across some modules---importantly between the Antennal Lobe and the Mushroom Body---occurs exclusively through the PNs, and the PNs have much higher levels of connectivity than the ORNs and KCs, which tend to be more module-specific. Like many artificial neural networks, the level of connectedness in the olfactory memory system increases at first---to facilitate comparisons and consolidation of information from different sources---and it reduces, as does the number of nodes, as the information flow approaches the output stage.

A third distinction in connectivity in biological and artificial networks can be gleaned from the literature and pertains to the level of specificity at which the connections operate. As \citet{Devineni2022} note in a useful survey, the variety of connections among \emph{Drosophila} neurons promotes versatility and makes the fruit fly responsive to a diversity of stimuli and contexts. The authors review an assortment of environmental and genetic interventions that researchers have performed to identify the neural pathways involved in such responses---as well as the evolutionary reasons why a stimulus such as the smell of carbon dioxide might elicit an attractive or aversive response depending on the situation. In typical artificial systems, as in the CNN structure in Figure \ref{fig:olfactory_cnn}, the level of connectedness is expressed by the values for the connection weights that are determined during the training process. The ability of such systems to exhibit responses that are customized to specific situations is limited by the depth of the network, with deeper structures capturing more complex interactions across regions of the input space. As an alternative to increasing depth, researchers often achieve some amount of context-sensitivity by introducing recurrence into artificial network structures, as discussed in Section \ref{Olfactory Coding and Memory}.

\subsection{Network Depth} \label{Depth}

\begin{figure}[ht]
    \centering
    \begin{subfigure}[ht]{0.22\textwidth}
        \centering
        \includegraphics[width=\textwidth,keepaspectratio]{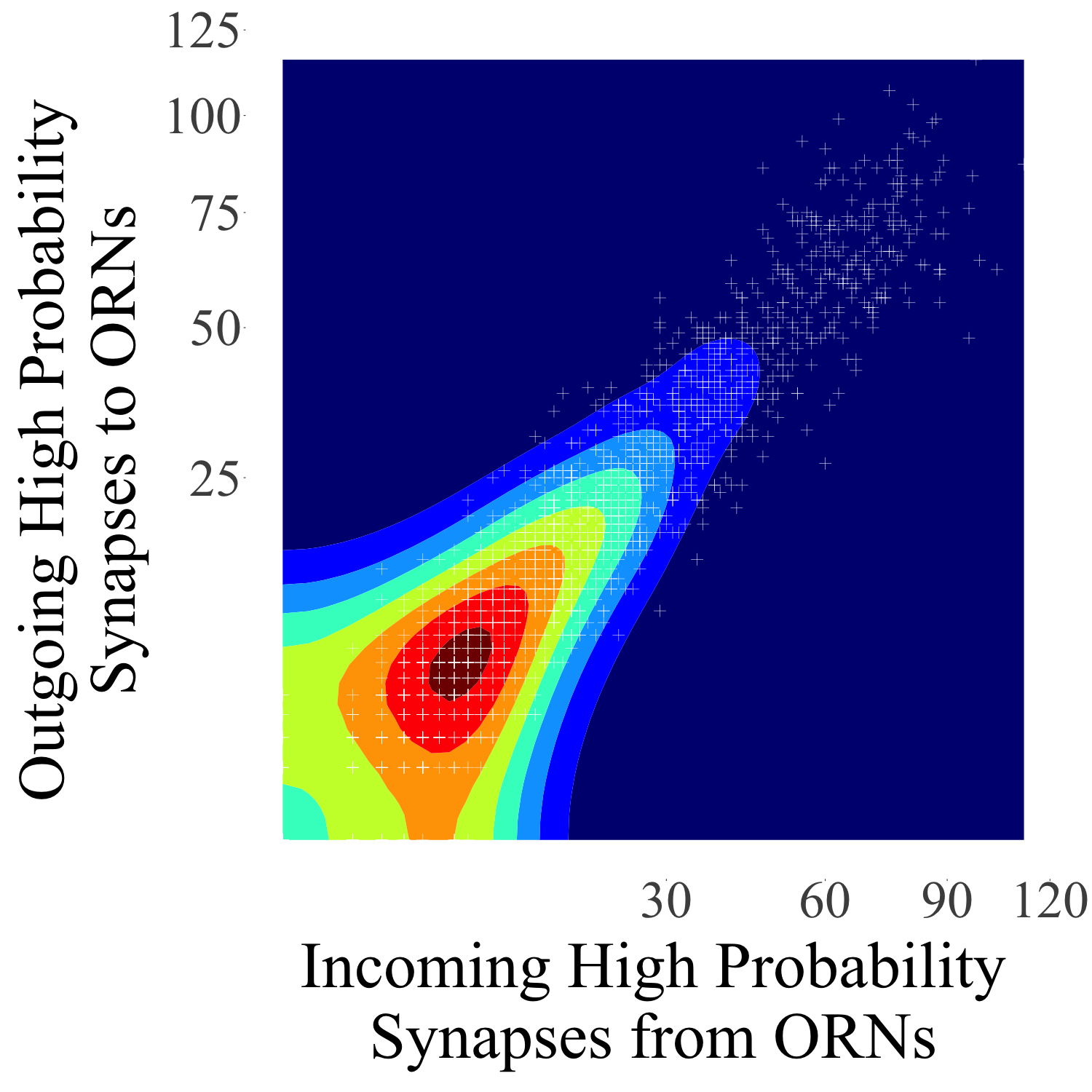} 
        \caption{from ORN vs. to ORN} \label{fig:fromORNtoORN}
    \end{subfigure}
    \hfill
    \begin{subfigure}[ht]{0.22\textwidth}
        \centering
        \includegraphics[width=\textwidth,keepaspectratio]{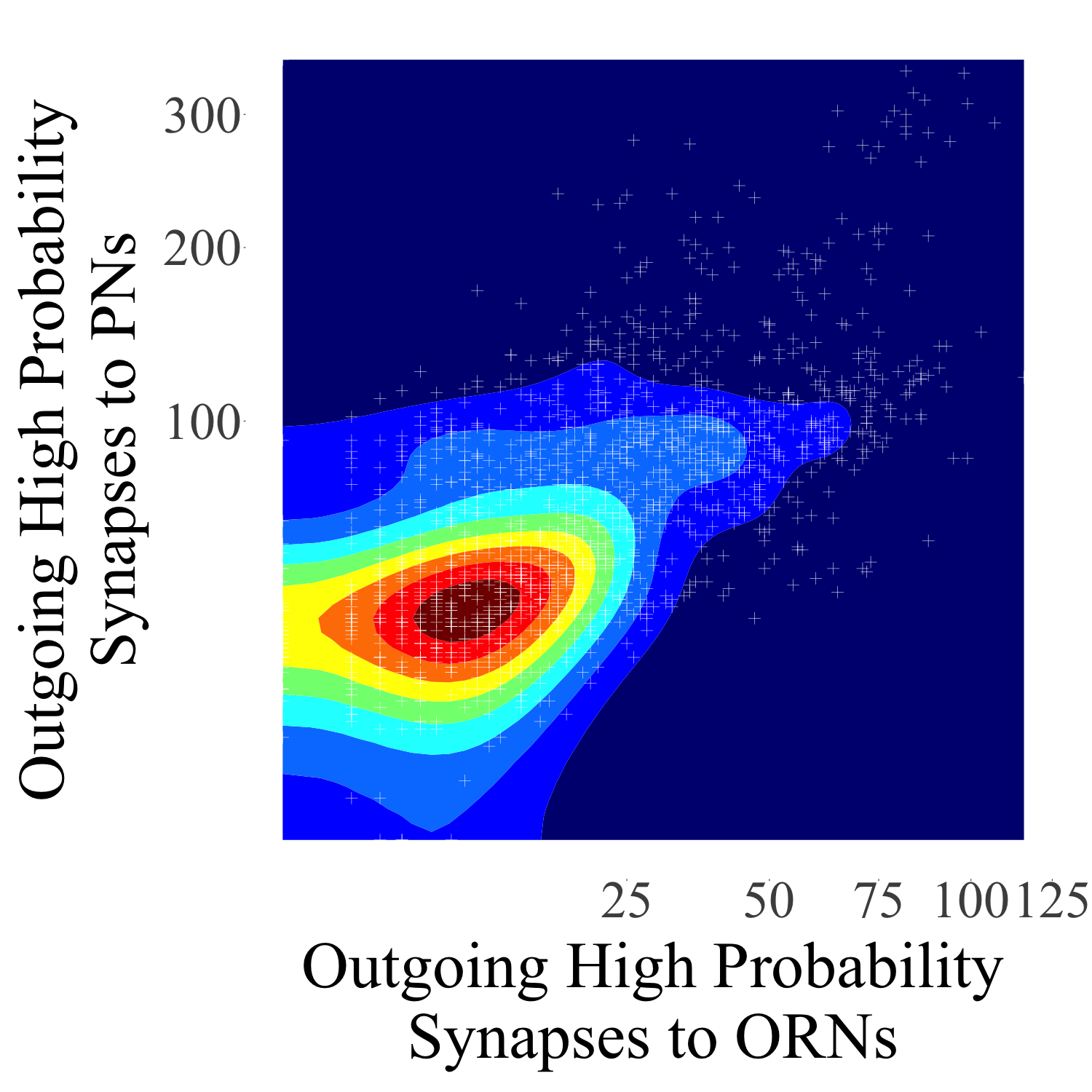} 
        \caption{to ORN vs. to PN} \label{fig:toORNtoPN}
    \end{subfigure}
\begin{minipage}{\linewidth}
\flushleft \tiny Notes: In each of these graphs, each observation represents one of the 1,474 olfactory neurons in the hemibrain data. Heatmaps are presented illustrating the joint densities, and the individual data points are overlaid as scatterplots. The values plotted along the vertical and horizontal axes are the total numbers of high probability synapses associated with that olfactory neuron that connect it to a neuron of a specific type. These numbers of high probability synapses between neurons of different types are the same as those totaled in Table \ref{table:syntype},, but here, they are broken out separately by neuron for the ORNs. In Figure \ref{fig:fromORNtoORN},, the vertical axis plots the numbers of high probability synapses that each ORN sends out to other ORNs, and the horizontal axis plots the numbers of high probability synapses that each ORN receives from other ORNs. The variation in the graph indicates that some ORNs are more connected than others, and the strong positive relationship indicates that there is not a tiered form of specialization in which some ORNs handle more raw information and others handle more processed information. In Figure \ref{fig:fromORNtoORN},, the vertical axis plots the number of high probability synapses that each ORN sends to PNs, and the horizontal axis plots the number of high probability synapses that each ORN sends to other ORNs. The positive relationship here indicates that there is also not a tiered form of specialization in which some ORNs primarily send signals to other ORNs while others primarly send processed output to PNs. The variation along both axes indicates that some ORNs transmit more information than others.
\end{minipage}
\vspace{0.10 in}
    \caption{High Probability Synapses to and from Olfactory Neurons} \label{fig:hpORN}
\end{figure}

\begin{figure*}[ht]
    \centering
    \begin{subfigure}[ht]{0.32\textwidth}
        \centering
        \includegraphics[width=\textwidth,keepaspectratio]{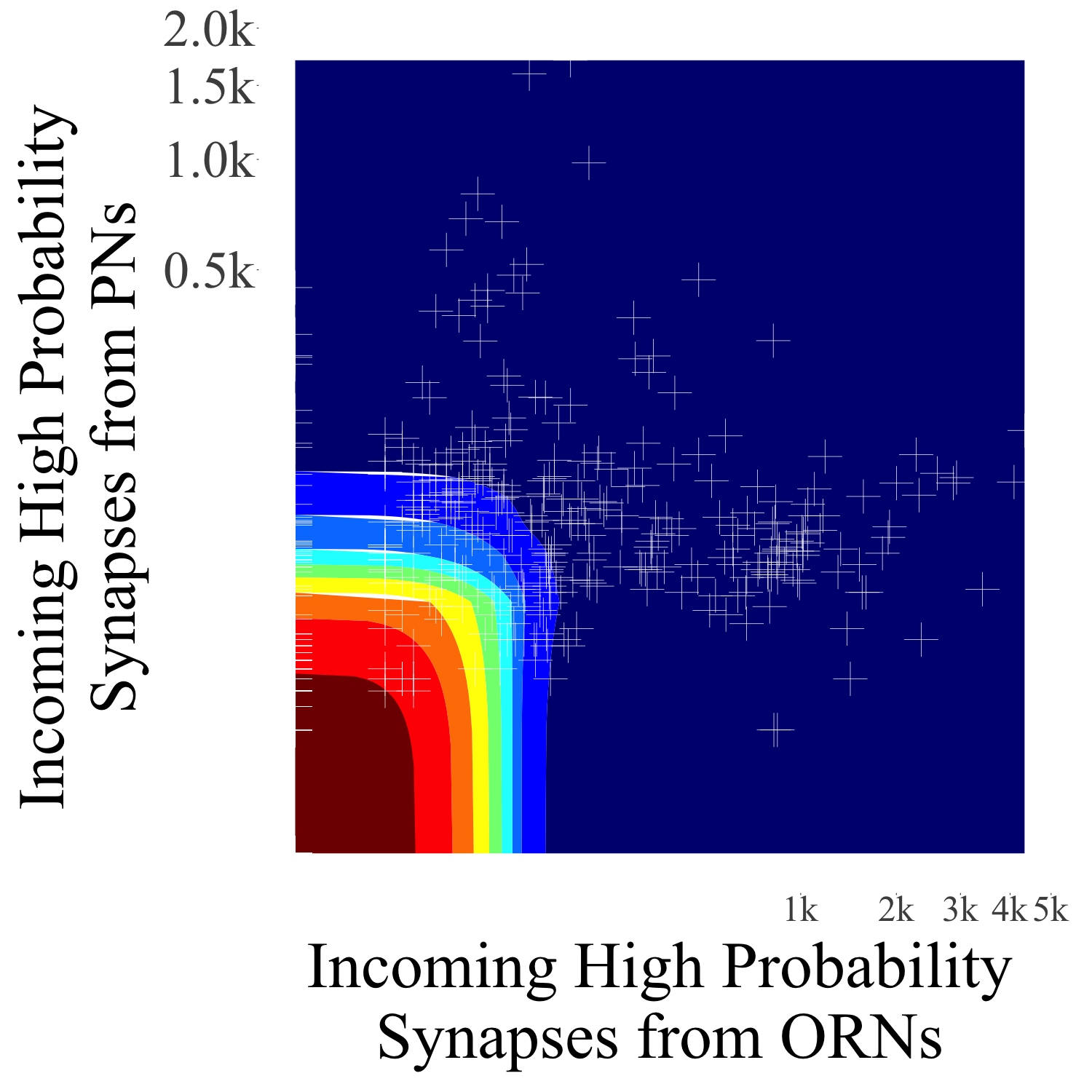} 
        \caption{from ORN vs. from PN} \label{fig:fromORNfromPN}
    \end{subfigure}
    \hfill
    \begin{subfigure}[ht]{0.32\textwidth}
        \centering
        \includegraphics[width=\textwidth,keepaspectratio]{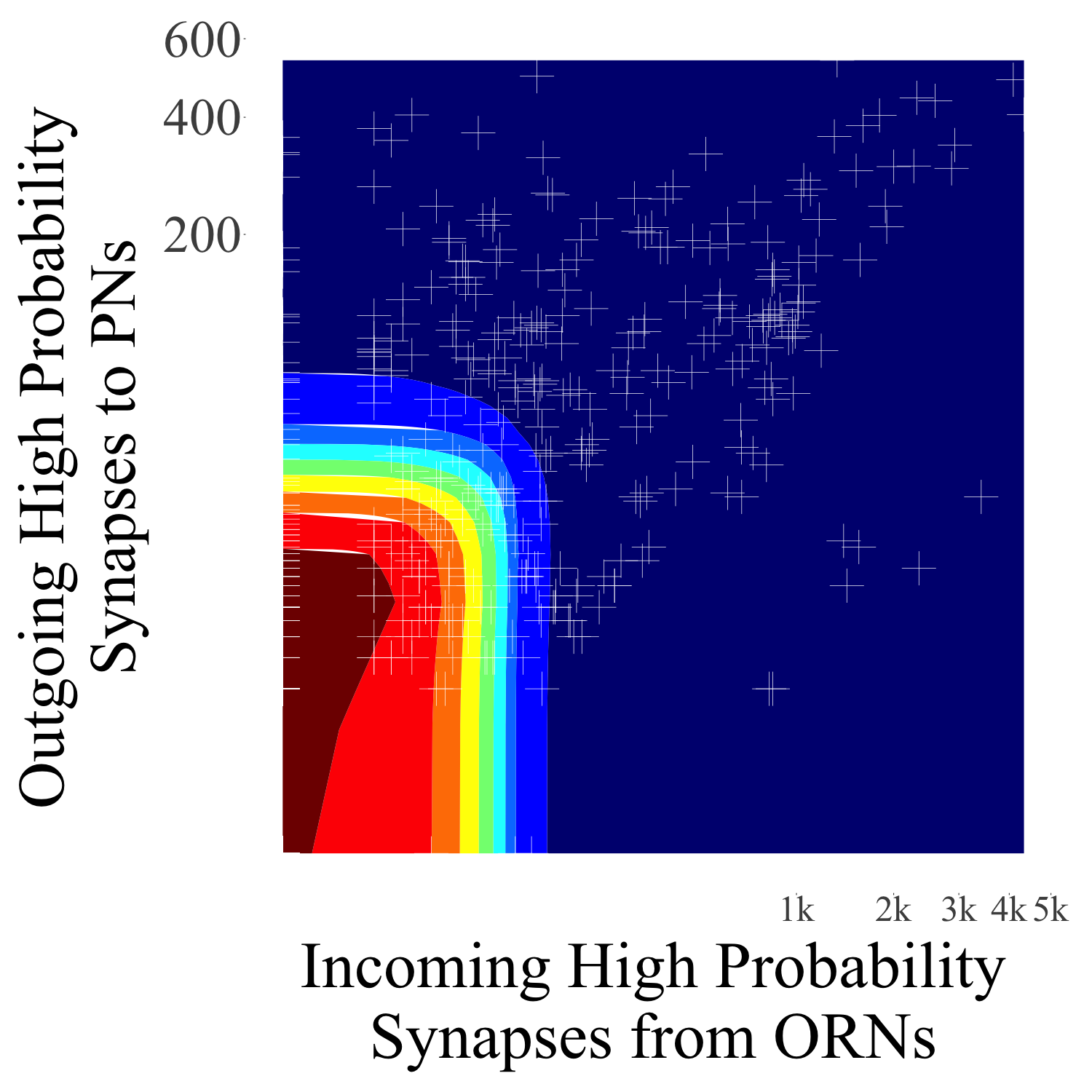} 
        \caption{from ORN vs. to PN} \label{fig:fromORNtoPN}
    \end{subfigure}
    \hfill
    \begin{subfigure}[ht]{0.32\textwidth}
        \centering
        \includegraphics[width=\textwidth,keepaspectratio]{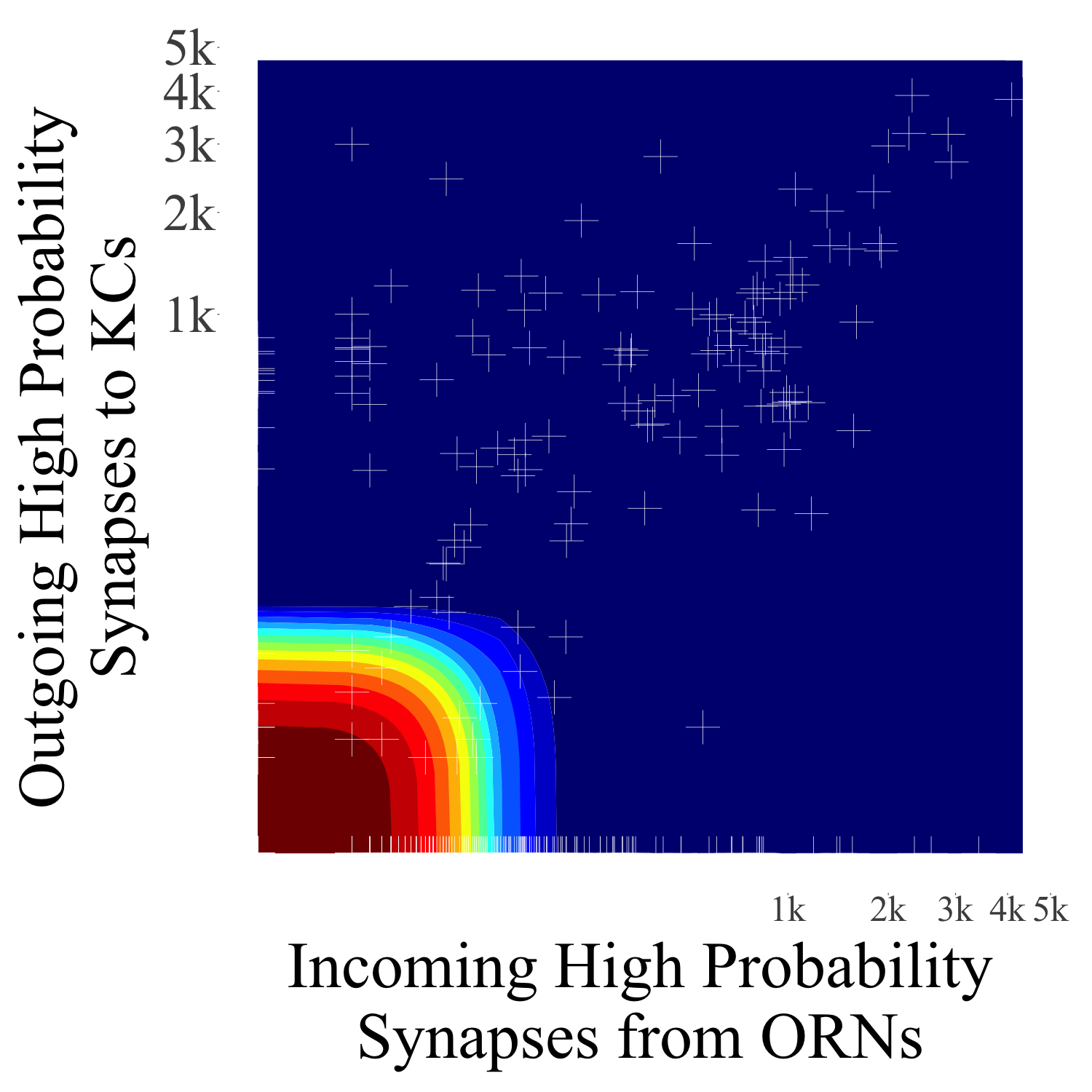} 
        \caption{from ORN vs. to KC} \label{fig:fromORNtoKC}
    \end{subfigure}
    \hfill
\begin{minipage}{\linewidth}
\flushleft \tiny Notes: The structure of these graphs is the same as in Figure \ref{fig:hpORN}, but in each of the figures, a single observation represents one of the 524 PNs in the hemibrain data. In all three graphs, the horizontal axis plots the numbers of high probability synapses that each of these PNs receives from ORNs. In Figure \ref{fig:fromORNfromPN}, the vertical axis plots the numbers of high probability synapses that each of these PNs receive from other PNs. There is some variation along this axis, indicating that some PNs receive more PN-related secondary input than others. This amount of secondary input is uncorrelated with the amount of primary input received from ORNs---and is not negatively correlated as one would expect to see in a deep tiered system of information flow. In Figure \ref{fig:fromORNtoPN}, the vertical axis plots the numbers of high probability synapses that each of the PNs sends to other PNs. The positive relationship indicates that PNs that receive more primary information from ORNs pass on more intermediate information to other PNs, but the relationship is not strong, and many of the PNs that transmit large amounts of information to other PNs have few high probability receiving synapses from ORNs. In Figure \ref{fig:fromORNtoKC}, the vertical axis plots the numbers of high probability synapses that each PN passes onto KCs in the mushroom body. While many of the PNs lie along the horizontal axis indicating no synapses to KCs, this relationship is generally positive, indicating that PNs that receive large amounts of primary olfactory information also transmit large amounts of information to KCs.
\end{minipage}
\vspace{0.10 in}
    \caption{High Probability Synapses to and from Projection Neurons} \label{fig:hpPN}
\end{figure*}

\begin{figure*}[ht]
    \centering
    \begin{subfigure}[ht]{0.32\textwidth}
        \centering
        \includegraphics[width=\textwidth,keepaspectratio]{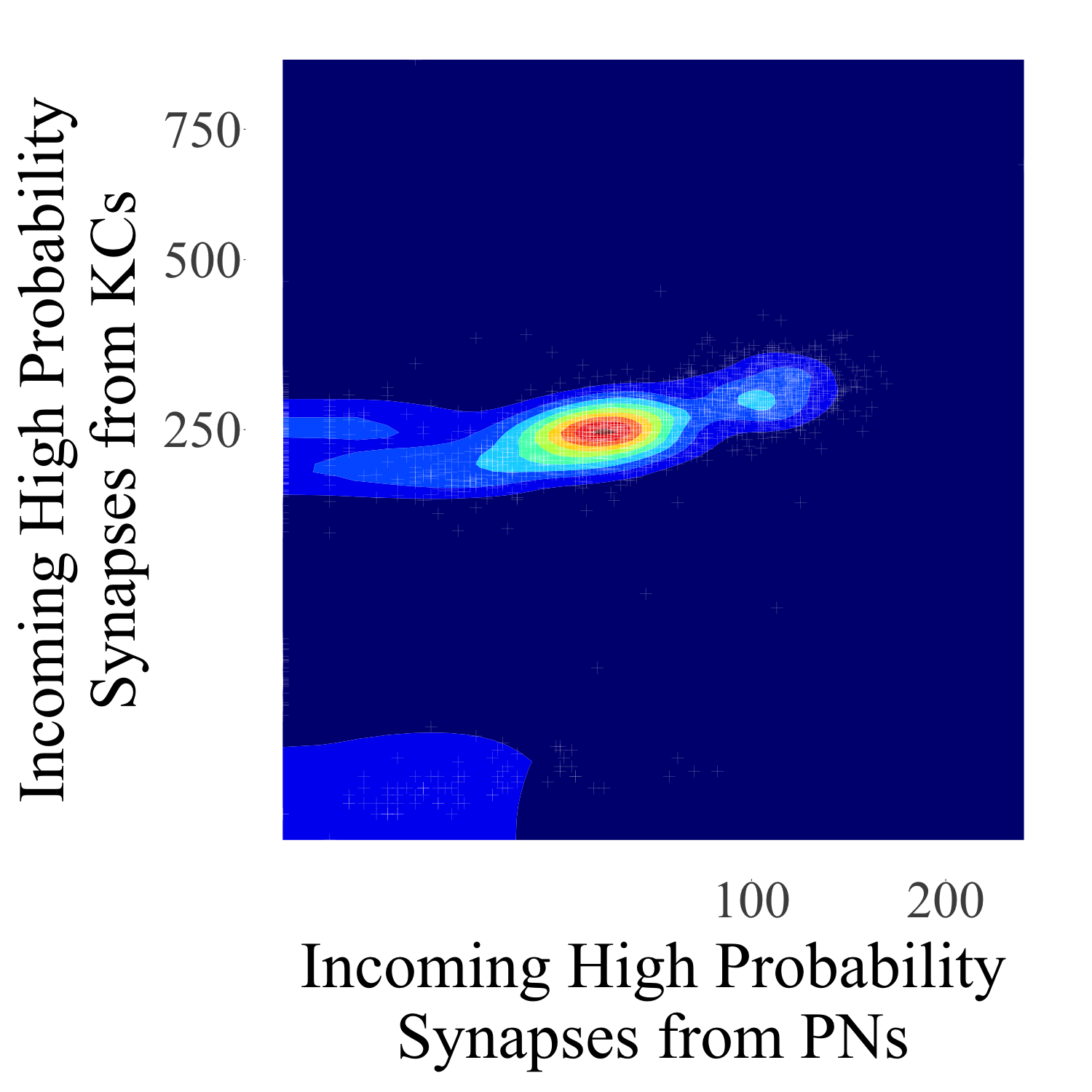} 
        \caption{from PN vs. from KC} \label{fig:fromPNfromKC}
    \end{subfigure}
    \hfill
    \begin{subfigure}[ht]{0.32\textwidth}
        \centering
        \includegraphics[width=\textwidth,keepaspectratio]{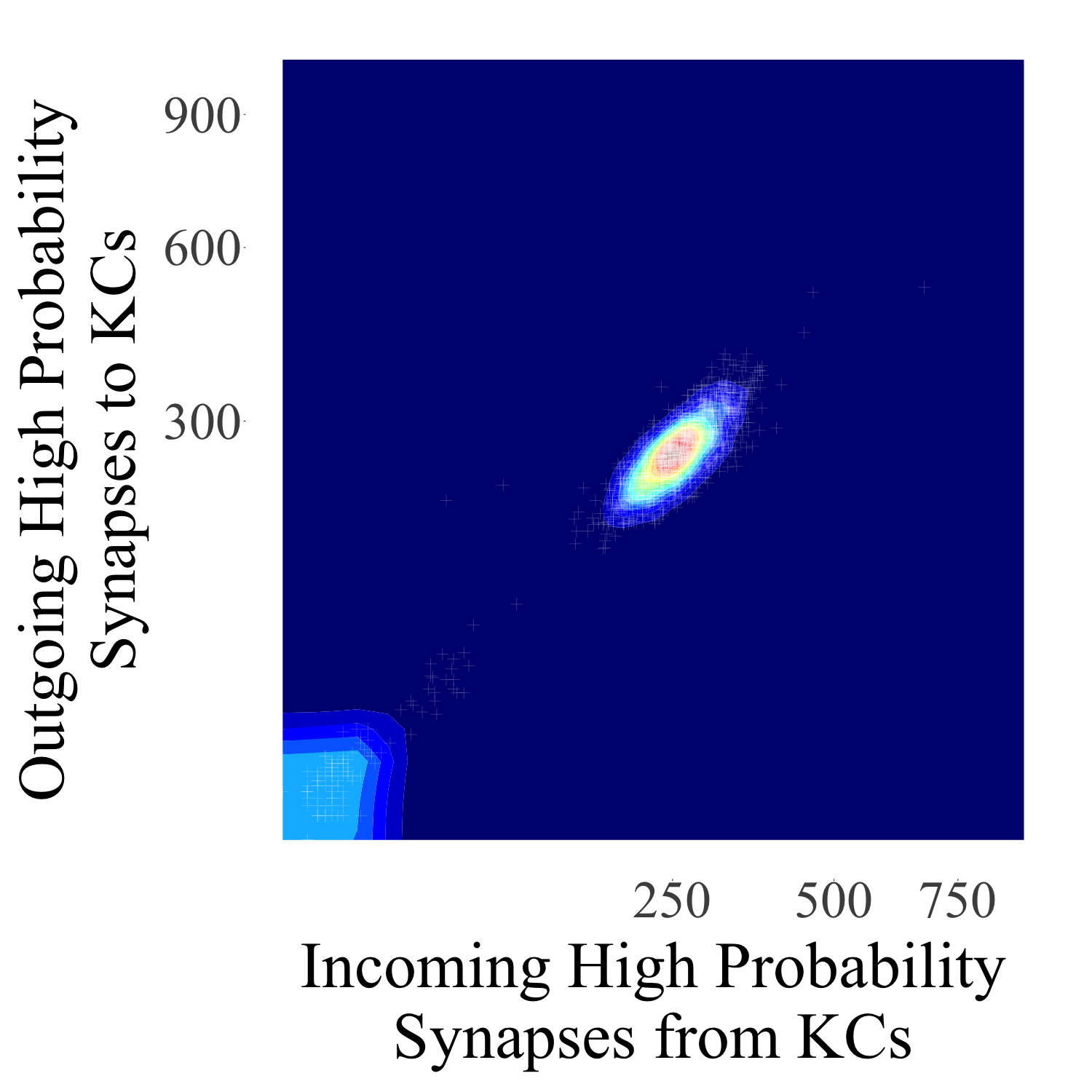} 
        \caption{from KC vs. to KC} \label{fig:fromKCtoKC}
    \end{subfigure}
    \hfill
    \begin{subfigure}[ht]{0.32\textwidth}
        \includegraphics[width=\textwidth,keepaspectratio]{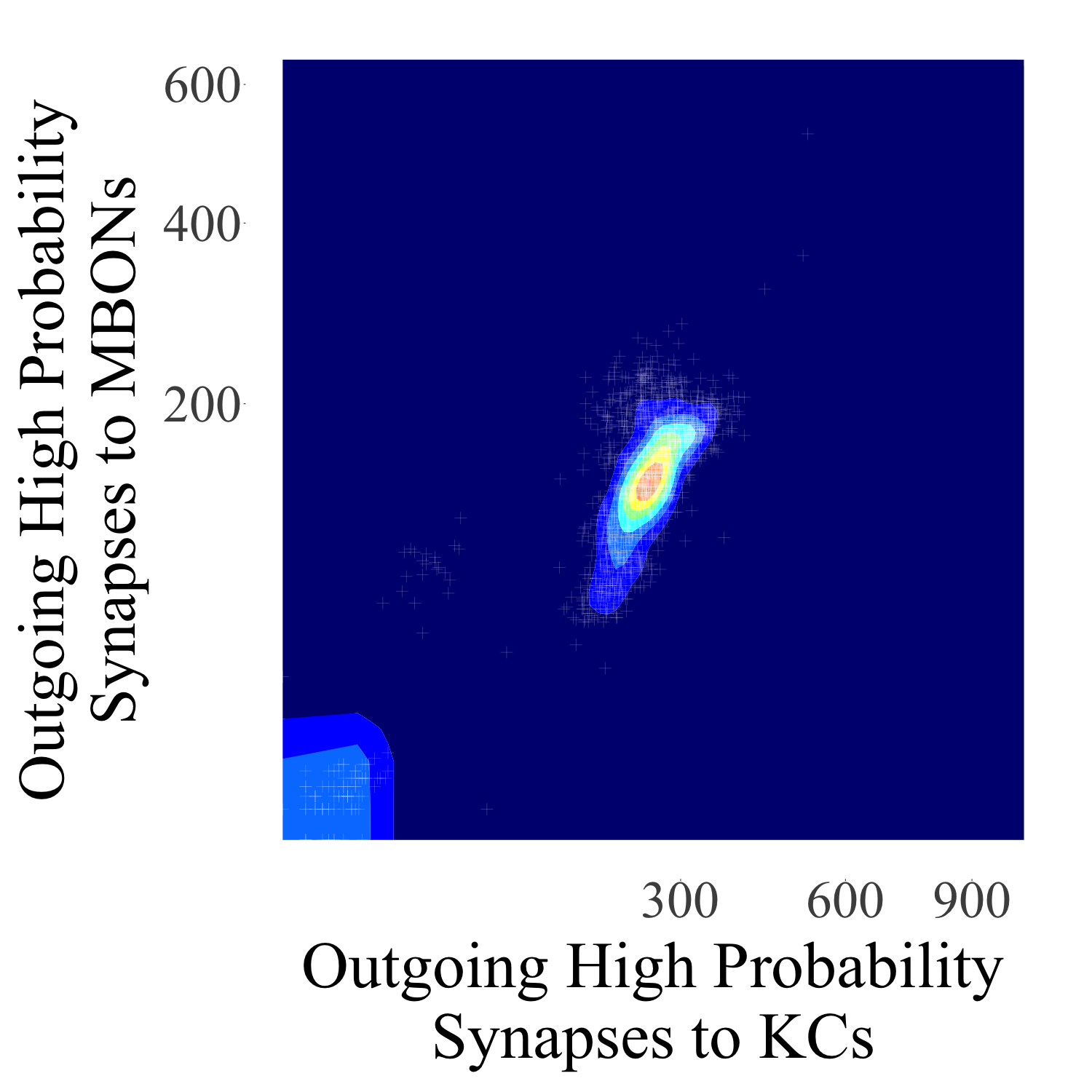} 
        \caption{to KC vs. to MBON} \label{fig:toKCtoMBON}
    \end{subfigure}
    \hfill
\begin{minipage}{\linewidth}
\flushleft \tiny Notes: The structure of these graphs is the same as in Figures \ref{fig:hpORN} and \ref{fig:hpPN}, but each observation represents one of the 2,033 KCs in the hemibrain data. In Figure \ref{fig:fromPNfromKC},, the horizontal axis plots the number of high probability synapses that each KC receives from PNs, and the vertical axis plots the number that each KC receives from other KCs. The clusters at the bottom along the left-hand side indicate that some KCs are specialized and mostly receive signals from one of these two input types. The large cluster in the center of the graph shows that the KCs have mostly high numbers of high probability synapses from other KCs---indicating a large amount of intra-layer communication among cells of this type---while the amount of input received from PNs is more varied. In Figure \ref{fig:fromKCtoKC},, the horizontal axis plots the number of high probability synapses that each KC receives from other KCs, and the vertical axis plots the number that each KC sends to other KCs. The numbers are highly positively correlated and exhibit little variance. As with the ORNs, there is not a tiered structure in which some cells receive primary information and others receive secondary information. Next, in Figure \ref{fig:toKCtoMBON} the numbers of high probability synapses received from other KCs are shown along the horizontal axis, and the numbers of outgoing high probability synapses to MBONs are plotted along the vertical axis. A small population of KCs in the lower left of the graph show low rates of output of either kind, though the large mass of KCs in the center exhibits consistently high levels of output to other KCs and some variation in how much output is transmitted to MBONs. Hence, the majority of KCs have common levels of intra-layer connectivity, although they vary in their amounts of signals sent outside the system through MBONs.
\end{minipage}
\vspace{0.10 in}
    \caption{High Probability Synapses to and from Kenyon Cells} \label{fig:hpKC}
\end{figure*}

To characterize in greater detail how these synaptic structures are laid out, Figure \ref{fig:hpORN} illustrates some key relationships between the numbers of high probability synapses of different types in the data in the earlier stages of the olfactory cascade. This figure for ORNs and the corresponding Figures \ref{fig:hpPN} and \ref{fig:hpKC} for PNs and KCs are formatted as joint density heatmaps showing the concentrations of neurons with different numbers of incoming and outgoing high probability synapses of different types. Each graph also presents the scatterplot from which its joint density is calculated; these points appear as white plus signs. Figure \ref{fig:fromORNtoORN} shows the relationship across Olfactory Neurons between the numbers of incoming high probability synapses received from other Olfactory Neurons and the numbers of outgoing high probability synapses to other Olfactory Neurons. Each plus sign in the scatter corresponds to a single ORN, with the heatmap illustrating concentrations of those ORNs. The vertical axis plots those neurons' numbers of outgoing high probability synapses whose target neurons are also ORNs, and the horizontal axis plots the neurons' numbers of incoming high probability synapses from source neurons that are also ORNs. The upward slope in the graph indicates that there is a strong positive correlation ($\rho = 0.932, t=98.8^{***}$)\footnote{This t-statistic is calculated using the \texttt{cor.test} function in R based on the assumption of independent observations. Asterisks indicate significance level (*** for 1\%, ** for 5\%, and * for 10\%).} between being highly connected to other ORNs as a source neuron and being highly connected to other ORNs as a target neuron. The ORNs shown in the upper right hand side of the graph show high levels of both types of connectivity, and the ORNs on the lower left show low levels of both types of connectivity. This positive relationship is consistent with a shallow, non-hierarchical structure for information processing among the ORNs. There are no ORNs in the upper left of the graph, who primarily send information to other ORNs but do not receive it, and there are also none in the lower right of the graph, who primarily receive information from other ORNs but do not send it. While some ORNs are more connected than others, the positive relationship between incoming connections from and outgoing connections to other ORNs does not support a model of layered processing of information within the set of ORNs. Next, Figure \ref{fig:toORNtoPN} shows the relationship across ORNs between the numbers of outgoing high probability synapses to other ORNs versus the numbers outgoing to PNs. The graph shows a positive correlation ($\rho = 0.609, t=29.4^{***}$), though not as strong as in Figure \ref{fig:fromORNtoORN}. This positive correlation indicates that ORNs are also not specialized in terms of who their targets are. The ORNs that send large amounts of output to PNs are the same ones that send large amounts of output to PNs. The two figures show that some ORNs have a hundred or more incoming ORN, outgoing ORN, or outgoing PN connections, while others have close to zero. Beyond the overall degree of connectedness, we see little evidence of specialization or a hierarchy in terms of the \emph{types} of connections observed across different ORNs.

Next, Figure \ref{fig:hpPN} shows the relationships across different types of connections further along in the olfactory cascade, among Projection Neurons. The structure of the figures is the same as in Figure \ref{fig:hpORN}, but within each panel, each observation is a PN. In all three panels, the number of incoming high probability synapses from ORNs is plotted along the horizonal axis, so that the amount of raw sensory input received is highest for the PNs on the right-hand side of the graph. Because the variable plotted along the horizontal axis is the same for the three panels, the horizontal placement of the 524 PNs is the same along all three panels, and only the vertical position---illustrating other forms of connectedness of the PNs---changes across the three. Due to the greater amounts of positive skew in these data and the concentrations of values around the origin, fourth-root scaling was applied to both axes, as compared with square-root scaling in Figures \ref{fig:hpORN} and \ref{fig:hpKC}, so that more of the data are visible on the graphs. This positive skew---with some very large values but a concentration around the origin---is the most prominent feature in each of the three figures. Overall, the three graphs are consistent with PNs exhibiting a wider variety of specialized roles than ORNs do in the olfactory cascade. This variety can be seen in the dispersion across highly positive values as well as the varying patterns of those values. In Figure \ref{fig:fromORNfromPN}, the vertical axis shows the number of incoming high probability synapses from other PNs. PNs with high values along the vertical axis receive large amounts of intermediate processed input. The correlation between the two variables is roughly zero ($\rho = 0.001, t=0.03$), indicating that there is little positive or negative relationship between the two types of inputs. The PNs receiving the highest amount of secondary input in the form of incoming high probability synapses from other PNs are comparable to other PNs in terms of the amount of primary input from ORNs that they receive. That is, there is not evidence of deep tiering like in artificial networks in which some PNs receive primary input and others receive secondary input---but there is some degree of secondary input processing that varies across PNs. In Figure \ref{fig:fromORNtoPN}, the vertical axis plots the number of outgoing high probability synapses to other PNs. There is a moderate positive correlation ($\rho = 0.540, t=14.7^{***}$) between the two variables; hence, PNs that receive greater amounts of raw sensory data tend to send greater amounts of data to other PNs, but there are exceptions to this tendency, and some PNs populate the upper left and lower right portions of the graph. In Figure \ref{fig:fromORNtoKC}, the vertical axis plots the number of outgoing high probability synapses that transmit information further along in the cascade to KCs. As with Figure \ref{fig:fromORNtoPN}, there is a moderate positive correlation ($\rho = 0.674, t=20.8^{***}$) in Figure \ref{fig:fromORNtoKC}. The PNs that receive large amounts of raw sensory input tend to be the ones transmitting large amounts of data to the KCs, but there are exceptions to the general tendency, with a concentration of PNs along the horizontal axis exhibiting no direct synaptic connections to KCs. 

\begin{figure}
\centering
    \begin{subfigure}[ht]{0.45\textwidth}
        \includegraphics[width=\textwidth,keepaspectratio]{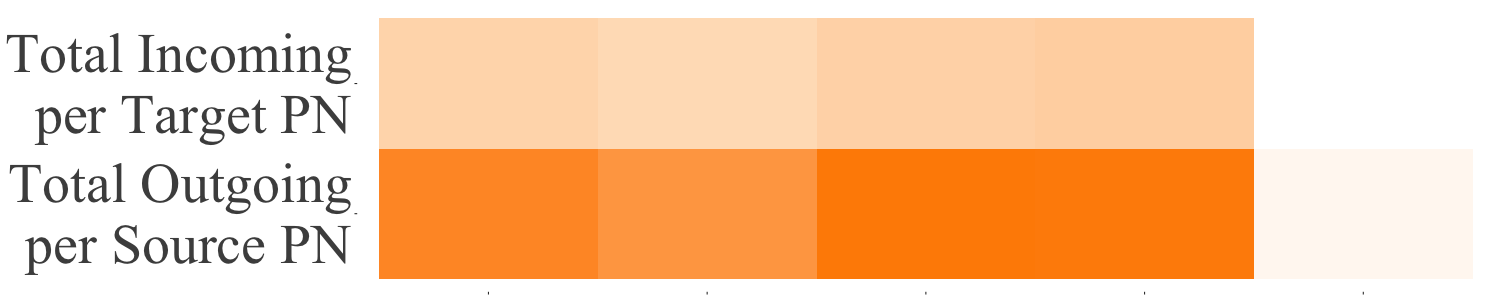} 
        \caption{Total} \label{fig:total}
    \end{subfigure}
    \begin{subfigure}[ht]{0.45\textwidth}
        \includegraphics[width=\textwidth,keepaspectratio]{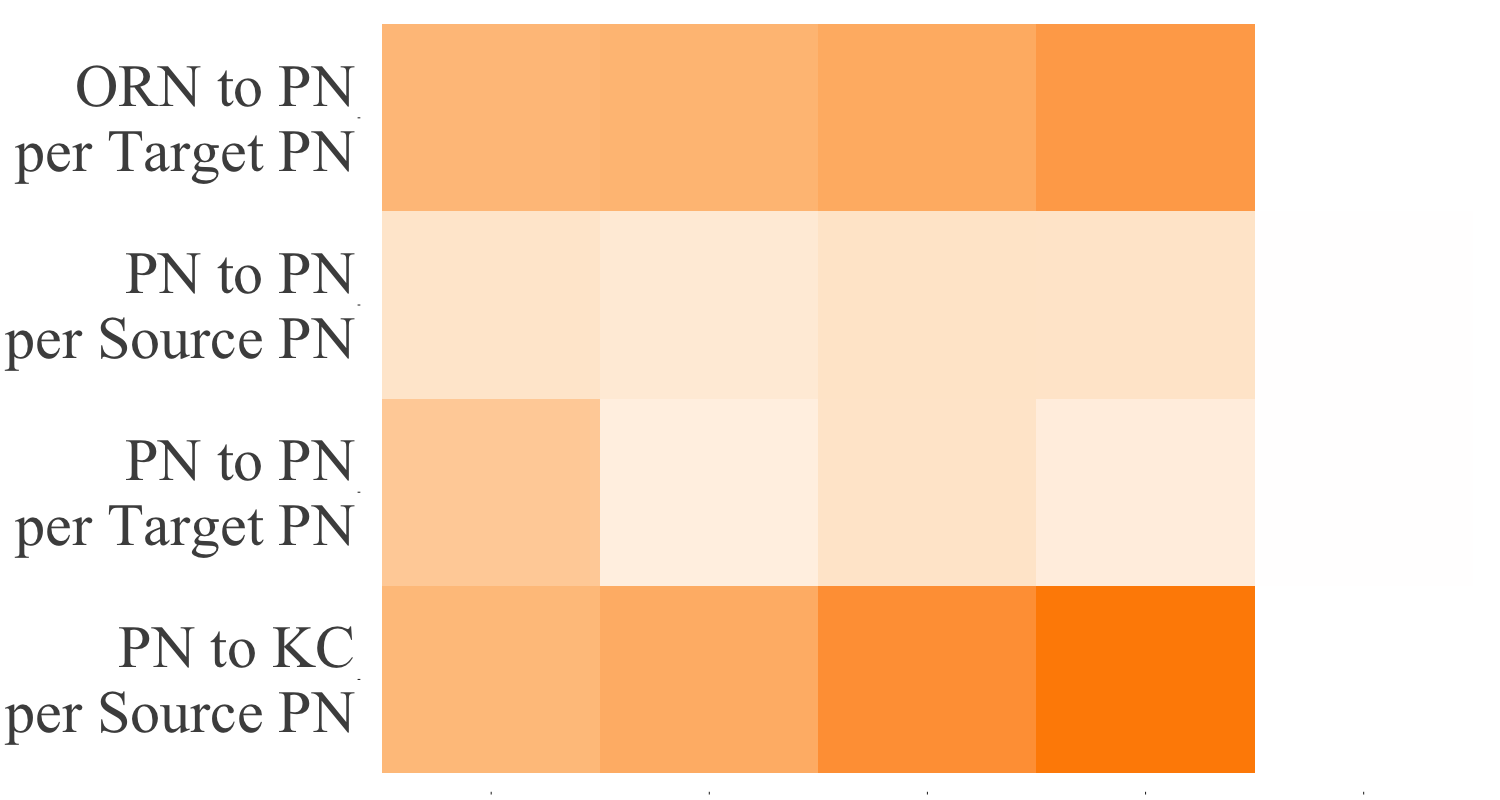} 
        \caption{By Source \& Target Type} \label{fig:breakdown}
    \end{subfigure}
    \begin{subfigure}[ht]{0.45\textwidth}
        \includegraphics[width=\textwidth,keepaspectratio]{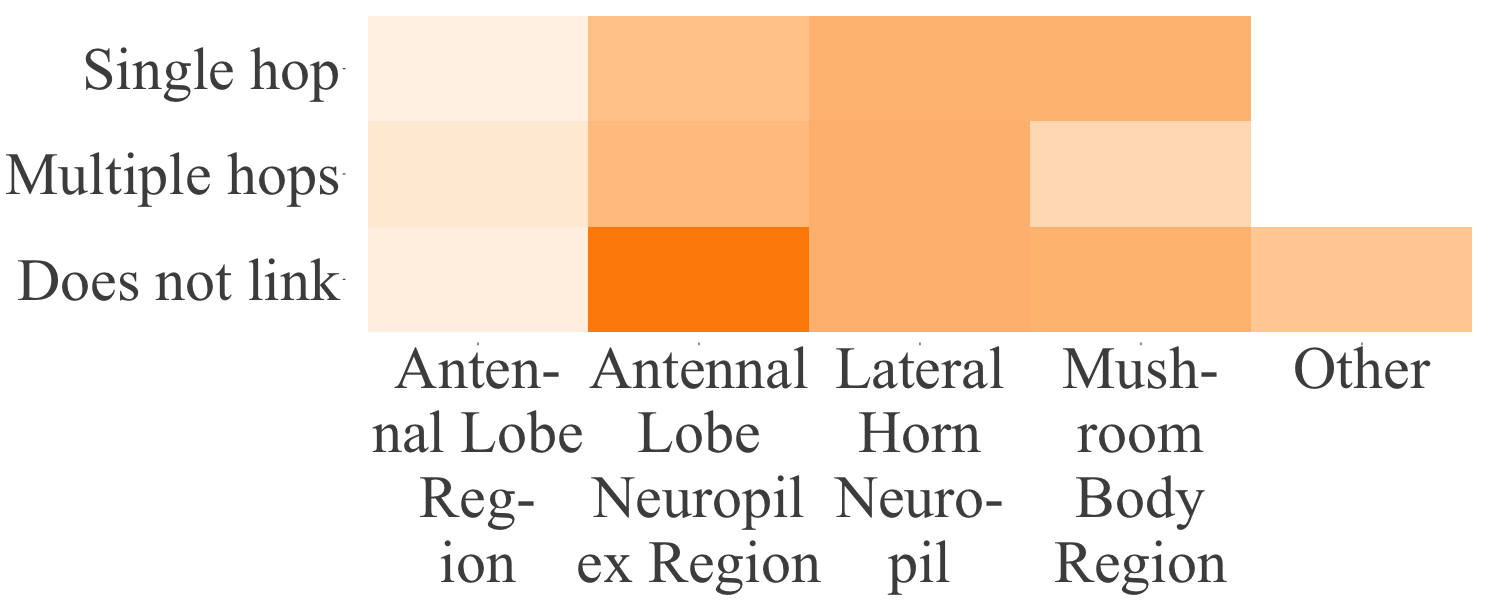} 
        \caption{By Role in the ORN-KC Channel} \label{fig:hops}
    \end{subfigure}
\begin{minipage}{\linewidth}
\flushleft \tiny Notes: These figures illustrate counts of Projection Neurons in the hemibrain dataset based on their locations, their levels of connectivity with other types of neurons, and their roles in the ORN-KC channel. The regional associations listed by the columns are the same as those in the columns of Table \ref{table:location}. Figures \ref{fig:total} and \ref{fig:breakdown} illustrate average numbers of incoming and outgoing high probability synapses per Projection Neuron, separately by the type of the sending or receiving neuron and the associated region. The average incoming and outgoing totals shown in Figure \ref{fig:total} are the same as the incoming and outgoing totals for PNs by location in panel B of Table \ref{table:spn}. The breakdown in Figure \ref{fig:breakdown} shows the portions of incoming that are received from ORNs and from PNs and the portions of outgoing that are sent to PNs and to KCs. The three rows of categories in Figure \ref{fig:hops} describe the manner in which a PN conveys information from ORNs to KCs. For a PN to be counted in the ``single hop'' row, it must have PN at least five incoming high probability synapses from a single ORN and at least five outgoing high probability synapses to a single KC. for a PN to be counted in the ``multiple hops'' row, it must be part of some sequence of ORN, a chain of PNs, and then a KC, where there are at least five high probability synapses corresponding to each of those linkages. If a PN is not counted in either of those first two rows, then it is counted in the third ``Does not link'' row, indicating that it does not play a substantive role in the transmission of olfactory information to the mushroom body.
\end{minipage}
\vspace{0.10 in}
\caption{Counts of Projection Neurons by Connectivity and Role in ORN-KC Channel}
\label{fig:pnsyn}
\end{figure}

To explore this specialization in PNs further, Figure \ref{fig:pnsyn} shows the intensity and distribution of the connections in this ORN to KC channel by illustrating the counts of different types of high probability synapses per PN as they vary across regions and neuropils, where the locations as before have overlap. Figure \ref{fig:total} shows the PNs' total numbers of incoming and outgoing connections by location, and Figure \ref{fig:breakdown} shows counts of total numbers of high probability synapses involved in the four key roles in the cascade---incoming from ORNs to PNs, outgoing from PNs to other PNs, incoming from other PNs to PNs, and outgoing from PNs to KCs. As the two rows of Figure \ref{fig:total} illustrate, the typical PN has considerably more outgoing than incoming connections. As Figure \ref{fig:breakdown} illustrates, the greatest number of direct incoming high probability synapses from ORNs and direct outgoing high probability synapses to KCs can be seen among PNs in the Mushroom Body. The PNs that are farther along in the circuit---those in the Lateral Horn and Mushroom Body---have more outgoing high probability synapses to KCs and also more incoming high probability synapses from ORNs. Hence, there appears to be a pattern whereby some PNs are especially important in the transmission of information to KCs, and those PNs are more likely than others to have a direct channel from stimulus to coding. The numbers of incoming high probability synapses from ORNs and outgoing high probability synapses to KCs are both highest among PNs associated with the Mushroom Body and are second highest among PNs associated with the Lateral Horn. The differences in these two synapse counts between those PNs associated with either the Lateral Horn or Mushroom Body and those associated with neither are $198.8, t=5.46^{***}$ for incoming synapses from ORNs and $304.1, t=9.15^{***}$ for outgoing synapses to KCs. Hence, in addition to providing more direct output, the amount of direct input received from ORNs is higher for PNs with locations deeper in the olfactory cascade. Thus, the connectivity is greater at the memory steps, but much of that connectivity is still direct. As can be seen in the middle two rows of Figure \ref{fig:breakdown}, the amount of PN-to-PN communication is non-negligible, particularly in the Antennal Lobe region.

Figure \ref{fig:hops} further illustrates the different roles that the 524 PNs play in linking ORNs to KCs and how it varies across regions. The three rows of that figure differentiate PNs based upon whether they provide information flow between ORNs and KCs---either on their own or in conjunction with other PNs. For the purposes of the figure, a PN has a direct connection if it has at least five incoming high probability synapses from ORNs and has at least five outgoing high probability synapses to KCs. A PN is considered to have an indirect connection to an ORN input if it has at least five incoming high probability synapses from PNs that themselves have five or more direct or indirect incoming high probability synapses from ORNs, and it is considered to have an indirect connection to a KC output if it has at least five outgoing high probability synapses to other PNs that themselves have five or more direct or indirect outgoing high probability synapses to KCs. A threshold of five high probability synapses is used so as to focus the discussion on neural connections that are biologically meaningful. The first row, labeled ``Single hop,'' indicates PNs that have at least five direct incoming high probability synapses from ORNs and at least five direct outgoing high probability synapses to KCs. The second row, labeled, ``Multiple hops,'' indicates PNs that have a role in the ORN to KC channel, but one or both of the connections is indirect. The third row, labeled ``Does not link,'' indicates PNs that either do not receive input (direct or indirect) from ORNs or do not submit output (direct or indirect) to KCs. Overall, 309 of the 524 PNs (59.0\%) do not have meaningful direct or indirect roles in the ORN-KC channel, 97 (18.5\%) have direct links, and 118 (22.5\%) have indirect links. Among the 215 PNs that do play a meaningful role, the shortest path from ORN to KC involves an average of 2.00 PNs or ``hops.'' Hence, there are multi-hop connections between ORN and KC---which is consistent with layered processing of information---but the networks tend to be shallow. These ORN to KC links are spread across regions in the olfactory cascade. While some fall into the Antennal Lobe areas, all of the direct single-hop links from the ORN to the KC are among PNs associated with both the Lateral Horn and the Mushroom Body, and none fall into the ``Other'' category in which they are not in any of the regions relevant to the cascade.

Moving further along in the signal chain, Figure \ref{fig:hpKC} illustrates some patterns of synaptic connections that can be seen among Kenyon Cells within the Mushroom Body. The structure of the figures is similar to that used in the previous two figures, but each panel shows distributions for the 2,033 KC neurons. First, Figure \ref{fig:fromPNfromKC} illustrates how the KCs vary in terms of the input signals they receive. The horizonal axis shows each KC's number of high probability synapses from PNs, and the vertical axis shows each KC's number of high probability synapses from other KCs. A distinct amount of specialization is apparent in the graph. The bulk of the cases are bunched in the center, with moderate numbers of incoming connections from both PNs and KCs, but there is another cluster of KCs at the bottom of the figure receiving a small fraction of their input from other KCs and a greater fraction directly from PNs. Additionally, along the verticial axis can be seen another small cluster of KCs that do not receive any of their signal from PNs. Overall, there is a moderate positive correlation ($\rho = 0.546, t=29.3^{***}$) between the two types of inputs. The distinct patterns from the graph are consistent with there being a fair amount of specialization among the KCs in the types of inputs they receive, with a segment of the population receiving particularly low amounts of signal from other KCs. In the second panel, the vertical axis of Figure \ref{fig:fromKCtoKC} plots the number of outgoing high probability synapses to other KCs, and the horizontal axis plots the number of incoming high probability synapses from other KCs. The two numbers of high probability synapses are highly positive correlated ($\rho = 0.920, t=105.5^{***}$). Hence, there is no indication of tiered processing whereby some KCs transmit secondary information from other KCs and others receive it. The results from these two panels some KCs are highly connected to other KCs, both with incoming and outgoing connections, while others are less so. The final panel, Figure \ref{fig:toKCtoMBON} plots the relationship between two output types---with numbers of high probability syanpses to MBONs along the vertical axis and numbers of high probability synapses to KCs on the horizontal axis. There is a positive correlation of $\rho = 0.760, t=52.7^{***}$, and the results are generally in line with those seen in the other panels. Some KCs are more connected than others. In addition to the general upward trend in the graphs, there is a distinct segment of KCs that are less connected to other KCs and MBONs; however, those KCs still receive moderate amounts of input from PNs.

The heterogeneous levels of depth across the system contrast with typical artificial systems such as the one illustrated in Figure \ref{fig:olfactory_cnn}, in which all variables and all observations pass through the same number of network layers.\footnote{In variations on this standard treatment, some artificial networks include skip patterns that combine information that has been subjected to different amounts of processing, as in the Deep Layer Aggregation approach of \citet{Yu2018}, while other artificial approaches calibrate the amount of processing at the observation level depending on the difficulty of the problem \citep{Rohlfs2022, Yang2020b}.} An additional distinction between the biological system and the artificial system is the shallowness of the overall biological network, particularly at the ORN stage. Due to artificial nodes' lower expressive power, longer chains of such neurons with intricate patterns of interaction along the way are required to convey the same level of signal specificity that is achieved with relatively shallow sequences of neurons in the olfactory system.

\section{Coding and Memory} \label{Olfactory Coding and Memory}

Beyond those factors that are apparent through the descriptive analysis above, some key differences between biological and artificial systems of information processing can be understood through how two systems operate and how memories are stored---which has been explored in the empirical literature. In a typical artificial neural network, learned information is retained primarily through the values of weights connecting different neurons---with some additional influence of experience on hyperparameters and modeling decisions made during validation. This development of neural connections over the learning process also occurs in the olfactory cascade, but there is an important additional way in which accumulated knowledge is represented in that system: the storage of memory in Kenyon Cells (KCs) in the Mushroom Body. These distinctions between biological and artificial learning are summarized in Table \ref{tab:coding}, which supplements the discussion of structural differences from Section \ref{Olfactory Anatomy} and Table \ref{tab:features}.

\begin{table}[h!]
\renewcommand\thetable{6}
    \caption{Coding and Memory in \emph{Drosophila} and Artificial Networks}
    \label{tab:coding}
\begin{threeparttable}
\renewcommand{\arraystretch}{2}
\centering
\resizebox{0.45\textwidth}{!}{
\begin{tiny}
\begin{tabular}{p{0.05\textwidth}|p{0.15\textwidth}|p{0.15\textwidth}}
Process & Biological System & Artificial Systems \\ \hline
Memory Representation & Kenyon Cells (KCs) in the mushroom body map to specific categories of stimuli that are generally fixed. Over the fly's lifetime, it learns positive and negative associations with these categories, and these associations are stored in the KCs. & Historical experience is primarily retained through the estimated values for the node-to-node connection weights---with some additional influence of history on hyperparameters and other broader modeling decisions. \\ \hline
Learning & Innate responses and categories for learning generally fixed for an individual and developed over millennia of evolution. Given the segregation of stimuli into these categories, an organism can learn, unlearn, and relearn associations and responses over its lifetime. & Relationships between inputs and outputs are learned by calibrating the weights of node-to-node connections to minimize errors in the training data. Broader modeling decisions such as the network structure are selected to minimize errors in a separate validation dataset. \\ \hline
\end{tabular}
\end{tiny}}
\begin{tablenotes}
\begin{subtable}{1.1\linewidth}
\vspace{0.10 in}
\tiny
\item Notes: The stability of the representations of sensations in KCs and the learning of associations with those representations are discussed in \cite{Aso2014b, Aso2016, Sitaraman2015, Ueno2017, Vogt2014} and \cite{Vogt2016}. The evolutionary element of training and the importance of innate responses---and in particular the distinction between that biological approach and training strategies used by artificial systems---are discussed in \cite{Sinz2019} and \cite{Zador2019}. Additional details about the representation of experience and the learning process in artificial networks can be found in \cite{GoodfellowBengioCourville2017}. Further discussion of the anatomical differences between these systems is provided in Section \ref{Olfactory Anatomy} and Table \ref{tab:features}.
\vspace{0.10 in}
\end{subtable}
\end{tablenotes}
\end{threeparttable}
\end{table}

\subsection{Memory in \emph{Drosophila}}

Experimental evidence from the neuroscience literature indicates that the mapping of sensory input types to different KCs is largely fixed. Using a light-based gene suppression approach, \cite{Aso2014b} identify specific PN channels that send dopamine signals to certain types of KCs in specific compartments in the mushroom body---with different channels involved in the formation of positive associations, the formation of negative associations, or signals inducing sleep. \cite{Sitaraman2015} also find specific circuits in the mushroom body that are dedicated to sleep regulation. \cite{Aso2016} refine this understanding of the roles of specific dopamine receptors by directly stimulating different receptors around the time that the flies are exposed to olfactory and visual cues. They find that identical neural pathways transmit information involved in the coding of positive or negative associations and also in the formation or reduction of such responses---with the effect type modulated by the timing of the signal relative to the stimuli.

\citet{Vogt2014} find that experiments to create positive and negative visual associations in memory stimulate some of the same locations in the mushroom body as are stimulated in the olfactory experiments, suggesting that there is a shared architecture for encoding such associations with different types of sensory inputs. \cite{Ueno2017} find a specific form of plasticity in \emph{Drosophila}'s olfactory memory that is activated only in the case of multiple forms of stimulation---indicating that memory modification is a sophisticated process that consolidates information from different sources. \cite{Vogt2016} find, however, that some locations in the mushroom body are specific to the encoding of associations with visual stimuli.

The roles that different parts of the brain play in influencing olfactory behavior is to a certain extent context-dependent. \citet{Bracker2013} find that the mushroom body impacts CO$_2$ avoidance for starved flies but not for fed ones. They hypothesize that, while CO$_2$ avoidance is largely viewed to be an innate response regulated by the lateral horn, the starved flies' brains deprioritize that response in favor of food search. The authors conjecture that the lateral horn suffices to produce the simple innate response, while the mushroom body is required to perform the more complex task of adjusting priorities in response to context. \citet{Devineni2022} provide a useful survey of context-dependent responses of this nature.

Some studies in Computational Biology have worked to explain essential patterns that have been observed in the olfactory cascade---particularly the manner in which which data from the ORNs is transmitted by a relatively sparse set of PNs to code the memories among larger set of KCs in the Mushroom Body. \cite{LitwinKumar2017} interpret the levels of connectivity of Kenyon Cells in the \emph{Drosophila} Mushroom Body from an information theoretic perspective. The authors contend that the sparsity of connections from PNs to KCs may assist in the efficient representation of learned associations by the KCs. \cite{Pehlevan2017} propose that the \emph{Drosophila} olfactory cascade can be modeled as a $k$-means clustering approach to coding olfactory associations and that some of the predicted qualitative patterns from that model are borne out empirically. \cite{Dasgupta2017} note that some nuances in the organization of these pathways matter in the efficiency of coding. The authors find that olfactory receptors and the corresponding neural pathways are organized according to the similarity of odors, so that, when a fruit fly learns an association with a specific odor, this learning also impacts the pathways and responses for similar odors. They relate this process to an algorithm known as locality-sensitive hashing.

\subsection{Memory in Artificial Systems}

\begin{table}[h!]
\renewcommand\thetable{7}
    \caption{Approaches for Representing Memory in Artificial Neural Networks}
    \label{tab:artificial}
\begin{threeparttable}
\renewcommand{\arraystretch}{2}
\centering
\resizebox{0.48\textwidth}{!}{
\begin{tiny}
\begin{tabular}{p{0.07\textwidth}|p{0.10\textwidth}|p{0.07\textwidth}|p{0.18\textwidth}|p{0.10\textwidth}}
Storage Location & Analogous Biological Process & Memory Type & Description & Sources \\ \hline
\multirow{3}{*}{\shortstack{Numeric\\weights\\connecting\\nodes\\across\\layers}} & \multirow{3}{*}{\shortstack{Formation of neural\\pathways over the\\lifetime and\\evolutionary history}} & Backpropagation & Primary means by which intensity of internode connections is learned in artificial neural networks & \cite{Rumelhart1986, GoodfellowBengioCourville2017} \\ \cline{3-5}
 & & Elastic Weight Consolidation & With each task learned, a new global constraint is added to the optimization function to penalize parameter updates that cause unlearning of prior skills & \cite{Kirkpatrick2017,Liu2020} \\ \cline{3-5}
 & & Synaptic Intelligence & Each synapse retains a parameter value and an ``importance,'' measure, so that updating is more costly for synapse weights that the network identifies as important for prior tasks & \cite{Zenke2017} \\ \hline
 \multirow{3}{*}{\shortstack{Earlier var-\\iables in\\sequence of\\observation-\\specific\\data flow}} & \multirow{3}{*}{\shortstack{Context-switching\\and prioritization such\\as that performed b\\the mushroom body}} & Recurrent Neural Networks & Structure for handling sequential data in which features' impact depends upon prior feature values & \cite{DeMulder2014,GoodfellowBengioCourville2017,Williams1986} \\ \cline{3-5}
 & & Long Short-term Memory & Form of RNN in which impact of prior values is context-dependent & \cite{Hochreiter1997,Smagulova2019} \\ \hline
Numbers of nodes in different layers & Neurogenesis and apoptosis & Neurogenesis & Replication and termination of network nodes as a function of firing intensity & \cite{Pandit2020} \\ \hline
Network architecture shared across tasks & Prioritization and higher-order components such as the mushroom body interacting with modules specialized for specific tasks such as olfaction & Progressive and Multi-task Learning & Learned parameters or architecture from previously encountered cases is maintained by the network and is consulted when updating model parameters. Buffers and policy functions that are shared across tasks or processing steps and retain information that passes a relevancy test & \cite{Fayek2020,Kerg2020,Rusu2016b,Rusu2016a,Teh2017} \\ \hline
\end{tabular}
\end{tiny}}
\begin{tablenotes}
\begin{subtable}{0.7\linewidth}
\vspace{0.10 in}
\tiny
\item Notes: This table describes means through which some artificial networks store data from recent experience or from prior observations and contexts.
\vspace{0.10 in}
\end{subtable}
\end{tablenotes}
\end{threeparttable}
\end{table}

The mapping of specific combinations of sensory inputs into memory centers with mutable stored responses provides insights for the design of memory structures in artificial systems. Work to incorporate memory into artificial networks has concentrated largely on the weights and connections in the network. Table \ref{tab:artificial} summarizes some of the methods that have been used in the literature to aid artificial networks in the coding and retrieval of past experiences and the ways in which they relate to structures in biological brains. 

The first three approaches listed, Backpropagation, Elastic Weight Consolidation (EWC), and Synaptic Intelligence (SI), all impact the the strength of connections across nodes in the network---an artificial counterpart to the numbers of synapses between neurons illustrated in Figures \ref{fig:hpORN}, \ref{fig:hpPN}, and \ref{fig:hpKC}. Backpropagation refers to the traditional manner in which learning proceeds in artificial neural networks, through which connection weights are calibrated to maximize training accuracy. EWC and SI are two means of modifying the network's optimization function to reduce the speed at which these weights are updated in order to preserve the level of accuracy in previously learned tasks. In biological systems, the same architecture is generally used for multiple tasks, and such neural structures are known to be less responsive to anomalous cases than are typical artificial networks. In that sense, decelerating the updating process may make artificial networks more like their biological counterparts.

The second category of memory retention strategies listed in Table \ref{tab:artificial} involves the use of Recurrent Neural Networks (RNNs)---of which Long Short-term Memory networks are a subset---to handle data that are sequential or have a time component, as in the case of speech or text. Through these systems, the response to an immediate stimulus can depend upon values observed one or more steps earlier in the sequence. This sensitivity to recent events resembles the context-specific prioritization observed by \citet{Bracker2013} to occur in the \emph{Drosophila} mushroom body.

The next strategy, Neurogenesis, is a biologically-inspired approach applied in the context of a Spiking Neural Network to mimic the processes through which learning impacts the rates at which different types of neurons are created and destroyed. This phenomenon is known to occur to a certain extent in the olfactory cascade---with KCs replicating as the number of mappings of stimuli into learned responses increases or the mappings become more granular. Hence, in addition to retaining the effects of past experiences through the strengths of connections across nodes, some recent work explores automated ways through which prior learning can impact network structure.

The final forms of memory discussed in the last two rows most closely resemble coding. In ``progressive'' systems, network architecture accumulates as new tasks are learned. In other networks built to handle multiple tasks, separate task-specific structures are developed that consult and occasionally modify a central policy function or set of constraints. Architectures of this form resemble the modular structure of biological brains, with some specialized components like olfactory neurons and other components with centralized roles, such as the mushroom body. In a related approach to multi-task learning by \citet{Borthakur2019} that is something of a hybrid between a biological and an artificial system, the authors design a Spiking Neural Network to mimic the structure of a mammalian olfactory system, and with a fixed set of hyperparameters, they find that the network exhibits strong performance on a variety of classification tasks.

All of these artificial frameworks have analogues to important ways in which biological brains store information; nevertheless, evidence from the fruit fly suggests that, at least in this one case of olfactory sensation, a substantial amount of memory formation occurs through a coding process that is fundamentally different from these methods that have been modeled in artificial systems. A distinct module in the brain serves the role of a lookup table. The overall structure through which memories are stored is stable, and the sensitivity of the system to past experience is largely contained among the KCs in this module. The formal characterizations by \cite{LitwinKumar2017,Pehlevan2017}, and \cite{Dasgupta2017} provide a starting point for thinking through biological strategies for representing sensory information in memory and ways in which such representations could be operationalized in the context of artificial neural networks. For the purposes of the comparison in Table \ref{tab:coding}, it is important that a representation of this form exists in a separate memory module and that this representation is generally stable over the organism's lifetime.

In an artificial system, a memory-based structure of this form could occur at an early stage in the processing of input data---either immediately following the input feed or further downstream after the convolutional layer and possibly after one or more hidden layers. The memory unit could in principle appear later in the network, but the earlier the placement, the more costly processing the system can replace with information learned from prior experience.

At whichever stage the memory unit appears, each observation being analyzed represents a point in the hyperspace of possible input values. A memory unit could carve this hyperspace into categories in the same way that the KCs correspond to specific combinations of sensory input. The grouping of the input space into these categories would be a stable modeling decision, just as each KC corresponds to a relatively fixed set of input signals. For instance, the categorization might slice the input space into a grid or use a kmeans-based approach to select natural groupings, with the exact formulation selected during the modeling or validation stage.

To align with the manner in which data are handled in the biological setting, two networks with the same grouping of inputs into memory categories could potentially learn opposite lessons about those categories, depending upon their respective experiences. At the training stage, as the artificial system encounters cases within each of these categories of inputs, it must determine how cases in these different categories are handled. The routing and dispatching of input cases by the memory system could resemble threshold-based decision rules like those used in \citet{Rohlfs2022, Yang2020b}, in which cases that can be classified with a high degree of confidence using the low-cost approach (in this case the stored response from memory) do so, reserving high-cost analysis for more difficult cases. For each category of inputs, the memory unit retains a vector of the counts of different output classes observed. If a sufficiently high number of cases have not been observed in that category, then the memory system passes it on for further analysis. If a large number of cases have been previously observed and the probability for the best classification exceeds some threshold, then the system bypasses the analysis and returns that classification.

\section{Conclusion} \label{Olfactory Conclusion}

This study provides a background and descriptive analysis of insect memory and the coding of olfactory sensation in \emph{Drosophila}. Neuron- and synapse-level data are explored from a newly available public source that covers more than half of the \emph{Drosophila} inner brain, including much of the olfactory processing system. The literature on coding and retrieval of olfactory memory in \emph{Drosophila} is discussed, with emphasis on differences between that biological system and typical artificial neural networks.

It is shown that, in contrast to typical artificial neural networks, which tend to have large numbers of layers, olfactory sensory information in \emph{Drosophila} is processed into memory using wide and shallow layers of communication. At both the initial sensory stage in the antennal lobe and at the coding stage in memory in the mushroom body, neurons vary widely in terms of their levels of connectivity, but they do not process information in a hierarchical or tiered fashion. Additionally, and also in contrast to artificial networks, neurons of the same type and in the same processing layer are highly connected with one another, indicating that some degree of coordination occurs at both the sensory and the coding levels.

Another important distinction is highlighted based upon findings from other empirical papers on \emph{Drosophila} olfaction. Learning in artificial neural networks generally occurs through the training process in which the weights of connections between nodes are calibrated---and in a less direct way through the validation process in which hyperparmeters such as the numbers of nodes are determined. By contrast, one key form in which memory is stored in \emph{Drosophila} is through a dedicated memory module: Kenyon Cells in the mushroom body. The mapping of specific sensory inputs to locations within this memory module is generally stable. The learning process focuses on the assignment of positive and negative associations with different segments of the sensory input space. A sketch is provided of potential ways in which an artificial system might incorporate such architectural features.

\bibliographystyle{cas-model2-names}
\bibliography{memory}%

\end{document}